\documentclass[a4paper]{article}
\usepackage[T1]{fontenc}
\usepackage[utf8]{inputenc}
\usepackage{amsmath,amsthm,amsfonts,amssymb,euscript,stmaryrd,graphicx,appendix}
\usepackage{algorithm,algorithmic}
\usepackage{color}
\usepackage{fullpage}
\usepackage{superbox}
\usepackage[shortlabels]{enumitem}

\def\simdist{\stackrel{\mathcal{L}}{\sim}}

\author{Sylvain \textsc{Corlay}\footnote{Bloomberg Quant Research, 731 Lexington Avenue, New York, NY 10022, USA. scorlay@bloomberg.net}}

\title{B-spline techniques for volatility modeling}
\date{May 29, 2014}

\newtheorem{theo}{Theorem}[section]
\newtheorem{prop}[theo]{Proposition}

\newtheorem*{remark}{Remark}

\newtheorem{defi}{Definition}[section]

\newcommand{\1}{\textbf{1}}

\newcommand{\Q}{\mathbb{Q}}
\newcommand{\R}{\mathbb{R}}

\newcommand{\E}{\mathbb{E}}

\newcommand{\PP}{\mathbb{P}}

\newcommand{\supp}{\operatorname{supp}}

\newcommand{\sspan}{\operatorname{span}}

\def\keywordname{{\bf Keywords:}} 
\newcommand{\keywords}[1]{\par\addvspace\baselineskip\noindent\keywordname\enspace\ignorespaces#1}

\graphicspath{{./}{figures/}}
\begin{document}
\maketitle
\begin{abstract}
\par This paper is devoted to the application of B-splines to volatility modeling, specifically the calibration of the leverage function in stochastic local volatility models and the parameterization of an arbitrage-free implied volatility surface calibrated to sparse option data. We use an extension of classical B-splines obtained by including basis functions with infinite support. 
\par We first come back to the application of shape-constrained B-splines to the estimation of conditional expectations, not merely from a scatter plot but also from the given marginal distributions. An application is the Monte Carlo calibration of stochastic local volatility models by Markov projection. 
\par Then we present a new technique for the calibration of an implied volatility surface to sparse option data. We use a B-spline parameterization of the Radon-Nikodym derivative of the underlying's risk-neutral probability density with respect to a roughly calibrated base model. We show that this method provides smooth arbitrage-free implied volatility surfaces. 
\par Finally, we sketch a Galerkin method with B-spline finite elements to the solution of the partial differential equation satisfied by the Radon-Nikodym derivative.
\end{abstract}
\vspace{2mm}
\section*{Introduction}
\par This article is concerned with the calibration of volatility models to market option prices. We address the particle method for the calibration of the leverage function in stochastic local volatility models, the problem of fitting a smooth and arbitrage-free implied volatility surface to sparse option data, and the numerical treatment of the Kolmogorov forward equation.
\par We dedicate the first section to giving background on B-splines. The classical B-spline basis functions have compact support and practitioners usually handle extrapolation by adding external ``ghost knots'' with a certain multiplicity. We favor an alternative extrapolation scheme which consists of supplementing the basis with functions of infinite support as proposed in \cite{SchumakerSplineBook}. 
\par Section \ref{sec:multiple_regression} is devoted to the problem of estimating a conditional expectation from given bivariate data and the knowledge of the marginal distributions. We review the Bayesian interpretation of Tikhonov regularization and we address the problem of compatibility between the marginal distributions and the conditional expectation. We show that shape-constrained B-splines are well suited to the problem of accounting for these compatibility conditions. Section \ref{sec:shape_constraints} gives some general background on second-order cone programming and its application to shape-constrained B-splines. 
\par In Section \ref{sec:particle_method}, we take on the recently devised particle method for calibrating the leverage function in stochastic local volatility models, where each time step requires the estimation of a conditional expectation, with a complete knowledge of the marginal distributions. We show that using the constrained regression method presented in Section \ref{sec:multiple_regression} to account for the compatibility conditions significantly improves the accuracy of the method.
\par Section \ref{sec:arbitrage_free_interpolation} is devoted to the problem of calibrating of a smooth arbitrage-free implied volatility surface from sparse option data. Our method is based on a B-spline parameterization of the Radon-Nikodym derivative with respect to a prior density. It allows for beliefs on the asymptotics of the volatility surface to be accounted for through the choice of a base model. While most approaches proposed in the literature rely on some kind of general-purpose nonlinear optimizer such as the Levenberg-Marquardt algorithm. Finally, the problem of calibrating an arbitrage-free surface to market option prices can be formulated as a second-order cone program, which is solved efficiently using off-the-shelf software such as CVXOPT \cite{cvxopt}, Mosek \cite{mosek} or CPLEX \cite{ilogcplex}. 
\par Finally, in Section \ref{sec:kolmo_radon_nikodym} we look at discretization schemes of the Kolmogorov forward partial differential equation. We propose using a B-spline-based finite element space discretization of the forward PDE satisfied by the Radon-Nikodym derivative of the underlying's risk-neutral distribution with respect to a base model. This leads to the same kind of surface parameterization as we have considered earlier.
\vspace{2mm}
\par \noindent \textbf{Notation:} We use the following conventions: $\inf(\emptyset) = +\infty$ and $\sup(\emptyset) = -\infty$. $\mathcal{P}_n(\R)$ is the set of real polynomials of order $n$. If $X$ is a random variable on the probability space $(\Omega, \mathcal{A}, \PP)$, $\PP_X$ denotes its pushforward measure. Indexing of knots and B-spline basis functions start at $0$.
\section{Univariate B-splines and extrapolation}\label{sec:spline_section}
\subsection{B-splines of infinite support}
\par In this section, we present the extension of classical B-splines devised in \cite{SchumakerSplineBook} by Schumaker to include basis functions with infinite support. 
\begin{defi}[B-splines of infinite support, \cite{SchumakerSplineBook}]\label{def:bsplines}
Let $k$ be a nonnegative integer and $\Gamma := \left\{ \gamma_0 \leq \gamma_1 \leq \cdots \leq \gamma_{k-1} \right\}$ be a sorted collection of $k$ knots. (If $k = 0$, $\Gamma := \emptyset$.) Let $C_0$ and $C_1$ be two positive constants. For a nonnegative integer $n \leq k$, a B-spline of order $n$ associated with the knots $\Gamma$ is a function of the form $\sum\limits_{j=0}^{k + n} w_j b^\Gamma_{j,n}$, where the weights $(w_j)_{ 0 \leq j \leq k + n}$ are real numbers and the functions $\left(b^\Gamma_{j,n}\right)_{n \geq 0, \ 0 \leq j \leq k+n}$ are defined by
\begin{equation}\label{eq:bsplines_order0}
\left\{\begin{array}{l}
b^\Gamma_{0,0}(x) := \1_{(-\infty,\inf(\Gamma))}(x), \qquad b^\Gamma_{k,0} := \1_{[\sup(\Gamma),+\infty)}(x),\\[4mm]
\textnormal{and} \quad b^\Gamma_{j,0}(x) = \1_{[\gamma_{j-1},\gamma_j)}(x), \quad 1 \leq j \leq k-1,
\end{array}\right.
\end{equation}
and for $1 \leq n \leq k$, with the induction formula
\begin{equation}\label{eq:bsplines_induction}
\left\{\begin{array}{ll}
b^\Gamma_{j,n}(x) := \frac{(\gamma_j - x)}{C_0} b^\Gamma_{j,n-1}(x), & j =0,\\
b^\Gamma_{j,n}(x) := b^\Gamma_{j-1,n-1}(x) + \frac{(\gamma_j - x)}{C_0} b^\Gamma_{j,n-1}(x), & 1 \leq j < \min(n,k),\\
b^\Gamma_{j,n}(x) := \left\{\begin{array}{ll} b^\Gamma_{j-1,n-1}(x) + \frac{\gamma_j - x}{\gamma_j - \gamma_{j - n}} b^\Gamma_{j,n-1}(x), & \textnormal{if} \ k > n,\\
b^\Gamma_{j-1,n-1}(x) + b^\Gamma_{j,n-1}(x), & \textnormal{if} \ k = n,
\end{array}\right. & j = \min(n,k),\\
b^\Gamma_{j,n}(x) := \frac{x - \gamma_{j-n-1}}{\gamma_{j-1} - \gamma_{j-n-1}} b^\Gamma_{j-1, n-1}(x) + \frac{\gamma_j-x}{\gamma_j - \gamma_{j-n}} b^\Gamma_{j,n-1}(x), & \min(n,k)+1 \leq j < \max(n,k),\\
b^\Gamma_{j,n}(x) := \left\{\begin{array}{ll} \frac{x- \gamma_{j-n-1}}{\gamma_{j-1} - \gamma_{j-n-1}} b^\Gamma_{j-1,n-1}(x) + b^\Gamma_{j,n-1}(x), & \textnormal{if} \ k > n,\\
b^\Gamma_{j-1,n-1}(x) + b^\Gamma_{j,n-1}(x), & \textnormal{if} \ k = n,
\end{array}\right. & j = \max(n,k),\\
b^\Gamma_{j,n}(x) := \frac{(x - \gamma_{j-n-1})}{C_1} b^\Gamma_{j-1,n-1}(x) + b^\Gamma_{j,n-1}(x), & \max(n,k) +1 \leq j < k + n,\\
b^\Gamma_{j,n}(x) := \frac{(x - \gamma_{j-n-1})}{C_1} b^\Gamma_{j-1,n-1}(x), & j = k+n.
\end{array}\right.
\end{equation}
\par \noindent Regarding the terms $\frac{x - \gamma_{j-n-1}}{\gamma_{j-1} - \gamma_{j-n-1}}$ and $\frac{\gamma_j - x}{\gamma_j - \gamma_{j - n}}$, the convention when the denominator is equal to zero is to replace it by $1$ and $0$ respectively. 
\par \noindent The so-defined collection of functions $\left(b^\Gamma_{i,n}\right)_{0 \leq i < k + n + 1}$ for $0 \leq n \leq k$ are called the B-spline basis functions of order $n$. The B-splines of order $n$ form a vector space of dimension $k + n + 1$.
\end{defi}
\begin{defi}[B-splines of higher order]\label{def:bsplines_higher_order}
With the same notation as in Definition \ref{def:bsplines}, for a nonnegative integer $n > k$, a B-spline of order $n$ associated with the knots $\Gamma$ is a function of the form $\sum\limits_{j=0}^{k + n} w_j b^\Gamma_{j,n}$, where the weights $(w_j)_{ 0 \leq j \leq k + n}$ are real numbers and where the functions $\left(b^\Gamma_{j,n}\right)_{n > k, 0 \leq j \leq k+n}$ are defined by the induction formula
\begin{equation}\label{eq:bsplines_induction_high_order}
\left\{\begin{array}{ll}
b^\Gamma_{j,n}(x) := \frac{(\gamma_j - x)}{C_0} b^\Gamma_{j,n-1}(x), & j =0,\\
b^\Gamma_{j,n}(x) := b^\Gamma_{j-1,n-1}(x) + \frac{(\gamma_j - x)}{C_0} b^\Gamma_{j,n-1}(x), & 1 \leq j < \min(n,k),\\
\left( b^\Gamma_{j,n}(x) \right)_{\min(n,k) \leq j < \max(n,k) + 1} & \textnormal{any basis of} \ \mathcal{P}_{\max(n,k)-\min(n,k)}(\R) = \mathcal{P}_{n-k}(\R)\\
b^\Gamma_{j,n}(x) := \frac{(x - \gamma_{j-n-1})}{C_1} b^\Gamma_{j-1,n-1}(x) + b^\Gamma_{j,n-1}(x), & \max(n,k) +1 \leq j < k + n,\\
b^\Gamma_{j,n}(x) := \frac{(x - \gamma_{j-n-1})}{C_1} b^\Gamma_{j-1,n-1}(x), & j = k+n. 
\end{array}\right.
\end{equation}
The positive constants $C_0$ and $C_1$ are the same as in Definition \ref{def:bsplines}
\end{defi}
\begin{remark}[On the choice of the constants $C_0$ and $C_1$]
A desirable property for the B-spline basis functions is that if the collection of knots $\Gamma$ is affinely transformed, the corresponding B-splines are affinely transformed as well. In other words, $C_0$ and $C_1$ should scale with $\Gamma$. In our implementation, we used $C_0 = C_1 = \frac{\gamma_{k-1} - \gamma_0}{k - 1}$ if $\gamma_{k-1} > \gamma_0$ and $C_0 = C_1 = 1$ otherwise. 
\end{remark}
\par
\begin{figure}[!ht]
	\begin{minipage}[c]{0.45\linewidth}
	\includegraphics[height=5.5cm]{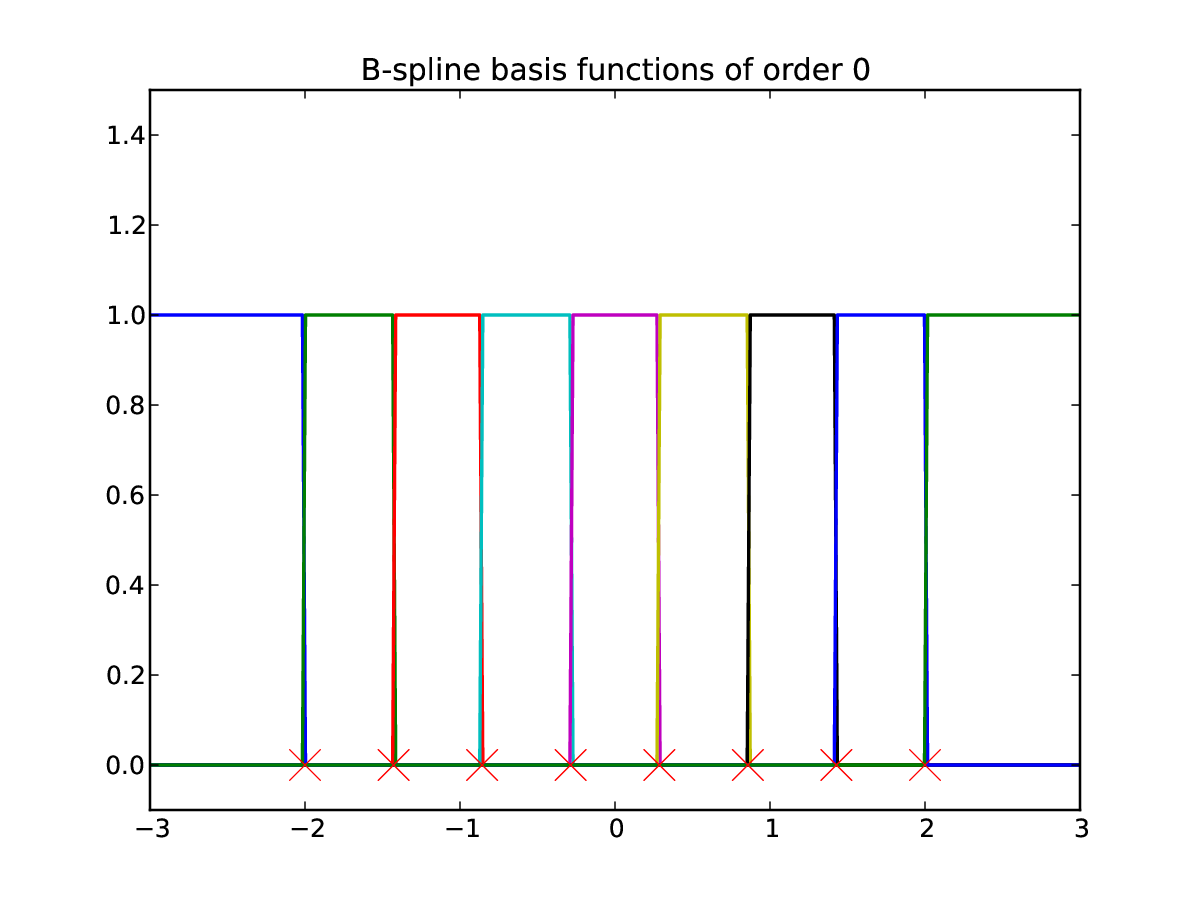}
	\end{minipage} \hfill
	\begin{minipage}[c]{0.45\linewidth}
	\includegraphics[height=5.5cm]{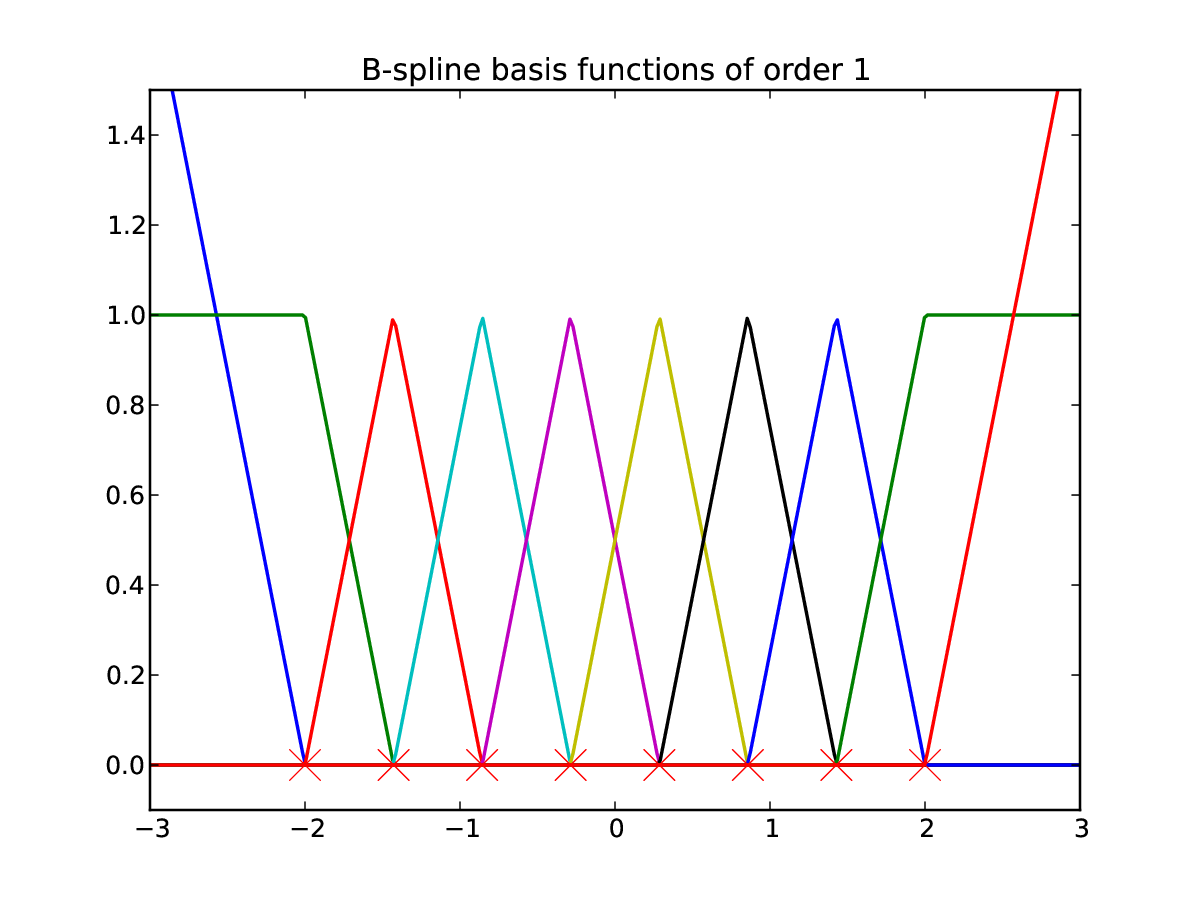}
	\end{minipage}\\
	\begin{minipage}[c]{0.45\linewidth}
	\includegraphics[height=5.5cm]{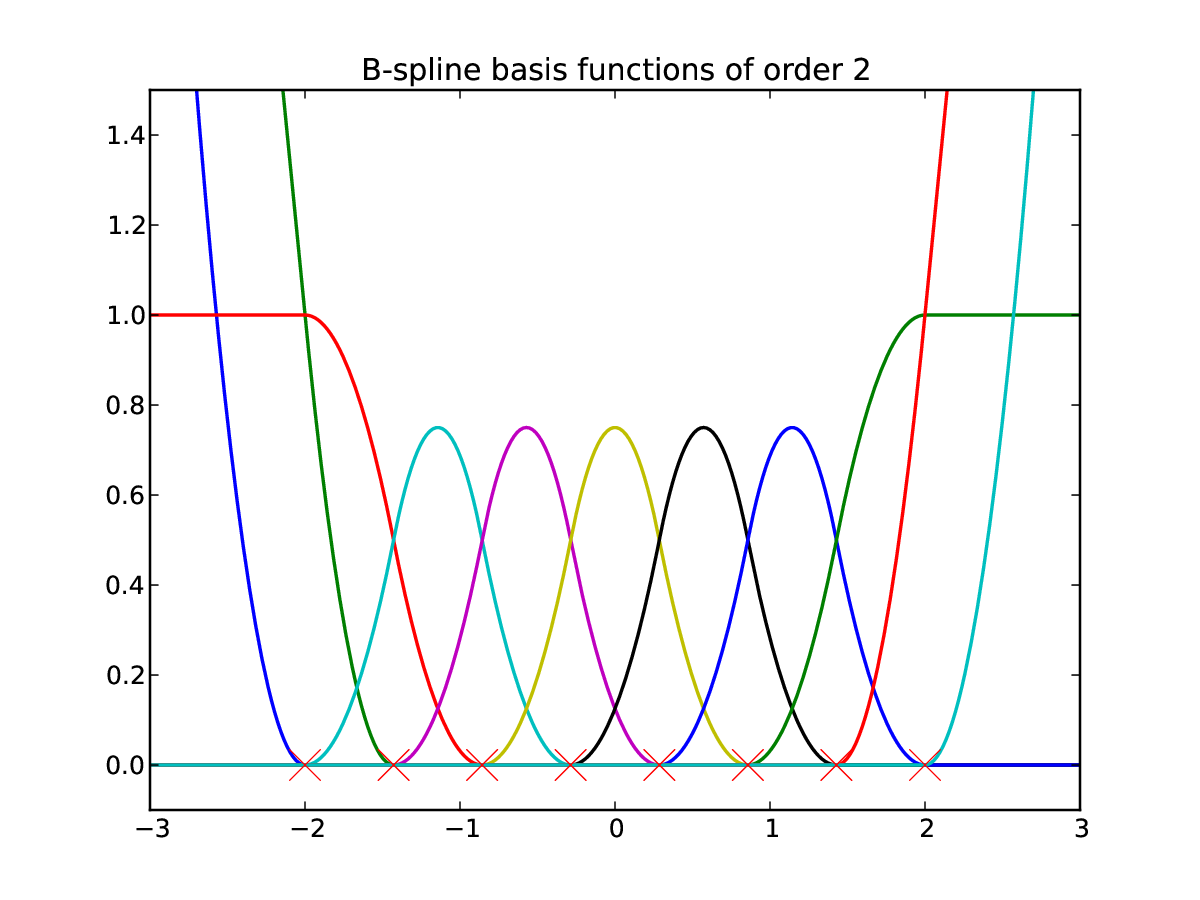}
	\end{minipage} \hfill
	\begin{minipage}[c]{0.45\linewidth}
	\includegraphics[height=5.5cm]{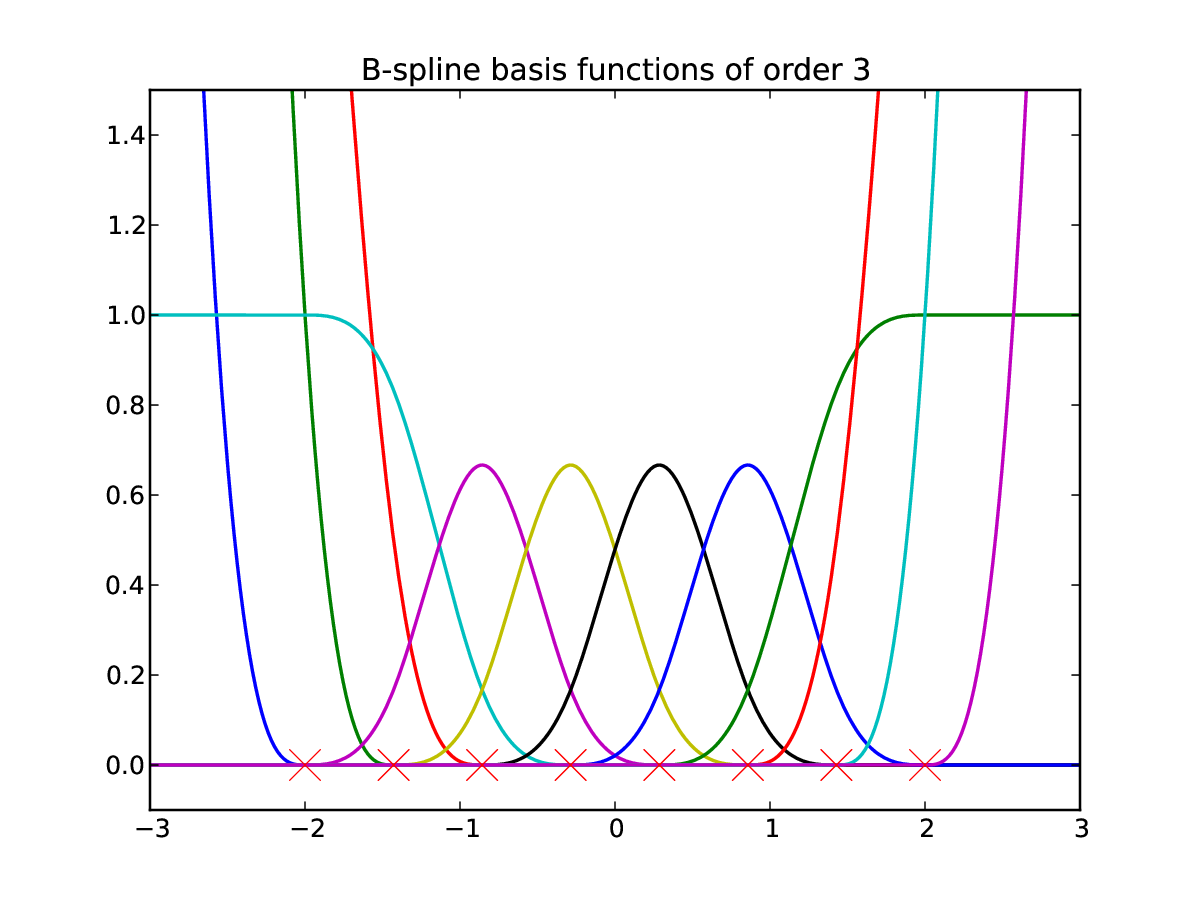}
	\end{minipage}
	\caption{The B-spline basis functions of order $0$, $1$, $2$ and $3$, corresponding to the same collection of $8$ knots.}
	\label{fig:b_splines_basis}
\end{figure}

 We have defined B-spline basis functions of arbitrary order associated with an arbitrary finite collection of knots. In Figure \ref{fig:b_splines_basis}, we display the B-spline basis functions of order $0$, $1$, $2$ and $3$, for the same collection of $8$ knots, where we have taken $C_0 = C_1 = \frac{\gamma_{k-1} - \gamma_0}{k - 1}$.
\vspace{2mm}
\begin{prop}[Properties of B-splines]
\par With the same notation as in Definition \ref{def:bsplines}:
\begin{itemize}
\item If $\gamma_0 < \gamma_1 < \cdots < \gamma_{k-1}$ (with strict inequalities), then the vector space spanned by $\left(b^\Gamma_{j,n}\right)_{0 \leq j \leq k+n}$ is the set of $C^{n-1}$ piecewise polynomial functions of order $n$ over $\R$ with breakpoints $\Gamma$. 
\item If $\gamma_{l-1} < \gamma_l = \cdots = \gamma_{l + m - 1} < \gamma_{l + m}$ for some $0 \leq l \leq k-1$ and $m\geq 1$, then the functions contained in $\sspan \left(b^\Gamma_{j,n}\right)_{0 \leq j \leq k+n}$ are piecewise polynomial functions of order $n$ and are only $C^{n - m}$ at $\gamma_l$. 
\end{itemize}
\end{prop}
\par \noindent In other words, the multiplicity of a knot diminishes the regularity of the spanned set of piecewise polynomial functions at the corresponding breakpoint.
\begin{remark}[Basis truncation]
\par When using B-splines for regression, a good way to avoid explosion of the extrapolation is to remove the basis functions of unbounded support that have a polynomial order strictly higher than $t$, $b^\Gamma_{0,n}, b^\Gamma_{1,n}, \cdots, b^\Gamma_{n - t - 1,n}$ and $b^\Gamma_{k + t + 1, n}, \cdots, b^\Gamma_{k + n, n}$.
The resulting vector space has dimension $k + 2 t - n + 1$. For $t=-1$, this reduces to the usual B-splines of compact support. 
\end{remark}
\par \noindent This illustrates one of the main advantages of this basis over the classical approach using ghost points, as limiting the extrapolation order simply amounts to truncating the basis rather than imposing linear equality constraints. 
\vspace{2mm}
\par \noindent The derivatives of these B-spline basis functions can be decomposed onto a B-spline basis of lower order. 
\begin{prop}[Differentiation of B-splines]\label{prop:b_spline_derivatives}
With the same notation, if $0 < n \leq k$,
\begin{equation}\label{eq:derivative_expr}
\left(b^\Gamma_{j,n}\right)' = \left\{\begin{array}{cl}
-\frac{n}{C_0}b^\Gamma_{j,n-1} & \textnormal{for} \ \leq j < \min(n,k), \\[2mm]
-\frac{n}{\gamma_n - \gamma_0} b^\Gamma_{n,n-1} & \textnormal{for} \ j = n, \\[2mm]
\frac{n}{\gamma_{j-1} - \gamma_{j-n-1}} b^\Gamma_{j-1,n-1} - \frac{n}{\gamma_j - \gamma_{j-n}} b^\Gamma_{j,n-1} & \textnormal{for} \ n+1 \leq j < k,\\[2mm]
\frac{n}{\gamma_{k-1} - \gamma_{k-n-1}} b^\Gamma_{k-1,n-1} & \textnormal{for} \ j = k, \\[2mm]
\frac{n}{C_1} b^\Gamma_{j-1,n-1} & \textnormal{for} \ \max(n,k)+1 \leq j < k + n + 1.
\end{array}\right.
\end{equation}
\end{prop}
\begin{prop}\label{prop:b_spline_derivatives_higher_order}
If we define the B-splines of order $-1$ by $b^\Gamma_{j,-1} := \delta_{x_j}$ for $j=0, \cdots, k-1$, where $\delta_x$ denotes the Dirac mass centered at $x$. If $k>0$ and $n=0$, we have
\begin{equation}\label{eq:higher_derivative_expr}
\left(b^\Gamma_{j,n}\right)' = \left\{\begin{array}{cl}
\textnormal{for} \ 0 & 0 \leq j < \min(n,k), \\[2mm]
\textnormal{for} \ -b^\Gamma_{n,n-1} & \textnormal{for} \ j = n, \\[2mm]
\textnormal{for} \ b^\Gamma_{j-1,n-1} - b^\Gamma_{j,n-1} & n+1 \leq j < k,\\[2mm]
\textnormal{for} \ b^\Gamma_{k-1,n-1} &\textnormal{for} \ j = k, \\[2mm]
\textnormal{for} \ 0 & \max(n,k)+1 \leq j < k + n + 1,
\end{array}\right.
\end{equation}
which can be seen as a limit case of Equation \eqref{eq:derivative_expr}.
\end{prop}
\par \noindent For $0 \leq p \leq n$, the decomposition of the $p$th derivative of B-spline basis functions of order $n$ onto the basis of order $n-p$ is obtained by iterating over this decomposition. 
\begin{remark}[Integration and inner products of B-splines]
\par Primitives and integrals of B-splines, as well as inner products of B-splines have closed-form expressions. An exact quadrature method is to use Gauss-Legendre points on each interval defined by the knots. A comprehensive study of methods to compute inner products of B-splines is carried out in \cite{VermeulentBartelsHeppler}. 
\end{remark}
\subsection{Evaluation and representation of B-splines}
\subsubsection*{The forward evaluation scheme for basis functions}
\par \noindent We can reformulate \eqref{eq:bsplines_induction} in a simpler way. Starting from $b^\Gamma_{j, n}(x) = 0$, we write for $0 < n \leq k$
\begin{equation}\label{eq:forward_method}
\begin{array}{ll}
\textnormal{For} \ 0 \leq j < \min(n,k) & \qquad b^\Gamma_{j, n}(x)\ +\!\! = \frac{\gamma_j - x}{C_0} b^\Gamma_{j, n-1}(x),\\
										& \qquad b^\Gamma_{j+1, n}(x)\ +\!\! = b^\Gamma_{j, n-1}(x),\\
\textnormal{For} \ \min(n,k) \leq j < \max(n,k) & \qquad b^\Gamma_{j, n}(x)\ +\!\! = \frac{\gamma_j - x}{\gamma_j - \gamma_{j - n}} b^\Gamma_{j, n-1}(x),\\
											   & \qquad b^\Gamma_{j+1, n}(x)\ +\!\! = \frac{x - \gamma_{j - n}}{\gamma_j - \gamma_{j - n}} b^\Gamma_{j, n-1}(x),\\
\textnormal{For} \ \max(n,k) \leq j < k + n & \qquad b^\Gamma_{j, n}(x)\ +\!\! = b^\Gamma_{j, n-1}(x),\\
											& \qquad b^\Gamma_{j+1, n}(x)\ +\!\! = \frac{x - \gamma_{j - n}}{C_1} b^\Gamma_{j, n-1}(x).
\end{array}
\end{equation}
\par \noindent This formulation is used to evaluate B-spline basis function in two ways:
\begin{enumerate}
\item The first and most natural approach is to use Formula \eqref{eq:forward_method} at each query point. 
\item The second method is to implement Formula \eqref{eq:forward_method} in terms of operations in the \emph{polynomial algebra}. With this pre-processing stage, we end up with a representation of the B-spline basis as the collection of their polynomial coefficients on each interval. 
\par The evaluation from the piecewise polynomial representation can be carried out using Horner's method, which is more efficient than recomputing the basis functions at new query points. 
\par Therefore the use of the second method, which involves a pre-processing stage, is beneficial if we evaluate the B-spline on a large number of points. The threshold for the number of evaluations is approximately equal to $n$ evaluations by interval. 
\end{enumerate}
\par \noindent In every case, one can use knowledge of the support of B-spline basis functions for their representation in memory and their evaluation. For a fixed $x \in \R$, if $0 \leq n \leq k$ there are $n+1$ B-spline basis functions of order $n$ that can be non-zero at $x$. More precisely, if $\gamma_i \leq x \leq \gamma_{i + 1}$ (with the conventions that $\gamma_{-1} = -\infty$ and $\gamma_k = +\infty$) the only B-spline basis functions that are not equal to zero are $b^\Gamma_{j,n}$ for $i \leq j \leq i + n$. 
\subsubsection*{The backward evaluation scheme}
\par \noindent Regarding the evaluation of a B-spline function $f = \sum \alpha_j b^\Gamma_{j, n}$, the natural and naive approach would be to use the forward evaluation scheme \eqref{eq:bsplines_induction_high_order} already presented for the basis functions and to compute the weighted sum. This is efficient if the B-spline basis functions have already been evaluated. However, if this is not the case, there is a more direct algorithm.
\par Indeed, using that the basis functions $b^\Gamma_{j, n}$ are decomposed onto the basis functions $b^\Gamma_{j-1, n-1}$ and $b^\Gamma_{j, n-1}$ (Equation \eqref{eq:bsplines_induction}), we can show that $f = \sum \alpha^{(1)}_j b^\Gamma_{j, n - 1}$ where the loadings $\alpha^{(1)}_j$ are piecewise polynomial of order $1$ and carry on with the decomposition of $b^\Gamma_{j, n - 1}$ onto a lower order basis. We find $f = \sum \alpha^{(i)}_j b^\Gamma_{j, n - i}$ where the loadings $\alpha^{(i)}_j$ are piecewise polynomial of order $i$. The algorithm stops when $i=n$ with the decomposition of $f$ onto the trivial basis $\left(b^\Gamma_{j,0}\right)_{0 \leq j \leq k}$. To get the loadings $\alpha^{(i+1)}_j$ from $\alpha^{(i)}_j$, we start from $ \alpha_j^{(i+1)}(x) = 0$ and write
\begin{equation}\label{eq:backward_method}
\begin{array}{ll}
\textnormal{For} \ 0 \leq j < \min(n - (i + 1),k) & \qquad \alpha_j^{(i+1)}(x) \ +\!\!= \alpha_j^{(i)}(x) \frac{\gamma_j -x}{C_0} + \alpha_{j+1}^{(i)}(x),\\
\textnormal{For} \ \min(n - (i + 1),k) \leq j < \max(n - (i + 1),k) & \qquad \alpha_j^{(i+1)}(x) \ +\!\!= \alpha_j^{(i)}(x) \frac{\gamma_j - x}{\gamma_j - \gamma_{j-n}} + \alpha_{j+1}^{(i)}(x) \frac{x - \gamma_{j-n}}{\gamma_j - \gamma_{j-n}},\\
\textnormal{For} \ \max(n - (i + 1),k) \leq j < k + n - (i + 1) & \qquad \alpha_j^{(i+1)}(x) \ +\!\!= \alpha_j^{(i)}(x) + \alpha_{j+1}^{(i)}(x) \frac{x - \gamma_{j-n}}{C_1}. 
\end{array}
\end{equation}
\par \noindent This method is called backward evaluation. The scheme was proposed in \cite[Chapter $5$]{SchumakerSplineBook} for the case of B-splines of compact support. It can be carried out in the polynomial algebra as well, to obtain a piecewise polynomial representation of $f$.
\begin{remark}
Backward and forward evaluation schemes can be used for the evaluation of derivatives of B-splines using Equation \eqref{eq:derivative_expr}. 
\end{remark}
\subsubsection*{The case of equally spaced knots}
\par A critical stage of all evaluation schemes is the localization of the query points in the knot vector. In the general case, this is done by bisection with $O\left(\log(k)\right)$ complexity. However, in the case where the knots are evenly spaced, this is reduced to an integer part computation. 
The case of equally spaced knots leads to further simplifications: all bounded spline basis functions have the same polynomial representation up to a parallel shift, and unbounded basis functions are symmetric. We can exploit these properties to save a significant amount of memory and computing. 
\section{Multiple regression and Bayesian considerations}\label{sec:multiple_regression}
\par In this section, we address the estimation of conditional expectations by multiple regression, with special attention paid to the case of the B-splines. We also recall the Bayesian theoretical foundation of Tikhonov regularization. 
\par Then we tackle the problem of estimating a conditional expectation, not merely from a scatter plot but also given the marginal distributions $\PP_X$ and $\PP_Y$. The regression problem must be constrained to account for these compatibility conditions. The problem can be formulated as a second-order cone program.
\par Eventually, we show that this technique can be used as a time-stepping scheme in the particle method proposed in \cite{GuyonLabordereParticular} for the calibration of stochastic local volatility models. 
\subsubsection*{Multiple regression as an approximation of conditional expectation}
\par Let $(\Omega, \mathcal{A}, \PP)$ be a probability space and $X$, $Y$ be two real random variables such as $Y \in L^2(\PP)$.
\vspace{2mm}
\begin{itemize}
\item $\E[Y|X]$ is the projection of $Y$ onto the vector space $\left\{f(X), f \in L^2(\PP_X) \right\}$, \textit{i.e.}, it is the solution of
$$
\min\limits_{f \in L^2\left(\PP_X\right)} \left\| Y - f(X) \right\|_2,
$$
\item while the multiple regression of $Y$ with respect to a finite collection $(f_i)_{i \in I} \in \left(L^2(\PP_X)\right)^I$ is the projection of $Y$ onto the subspace $\sspan\limits_{i \in I}\{f_i(X)\}$, \textit{i.e.}, it is the solution of
$$
\min\limits_{w \in \R^I} \left\| Y - \sum\limits_{i\in I} w_i f_i(X) \right\|_2.
$$
\end{itemize}
\vspace{2mm}
\par \noindent Hence, the larger the vector space $\sspan\limits_{i \in I} (f_i)$, the better the approximation of $\E[Y |X]$ by the multiple regression of $Y$ with respect to $(f_i(X))_{i \in I}$. 
\subsection{Regression of empirical distributions}
\par In practice, we usually only have a finite sample $(x_j, y_j)_{1 \leq j \leq N}$ of independent draws from $(X,Y)$. A common approach is then to approximate the multiple regression of $Y$ with respect to $(f_i(X))_{i \in I}$ by the regression of the corresponding empirical distributions. When doing so, enlarging the vector space onto which we project can be detrimental rather than beneficial. Indeed, performing a better regression of the empirical distribution does not mean that we get a better regression of the actual distribution of $X$ with respect to $Y$. This phenomenon, also called ``over-fitting'', occurs for example when using a very fine grid for piecewise linear regression. Certain practitioners refrain from using a fine grid because of it. This means that they do not believe in wiggly results, that is, they have a prior belief on the smoothness of the conditional expectation. 
\par Rather than refraining from refining the grid, another approach to the problem of over-fitting is the Bayesian approach, that is, to determine the most likely conditional expectation of $Y$ with respect to $X$ given the observed sample $(x_j, y_j)_{1 \leq j \leq N}$ and the prior distribution for the conditional expectation.
\subsubsection*{Bayesian foundations of Tikhonov regularization}
\par We now assume that $X$ and $\mathcal{E}$ are $L^2$ real random variables and $F$ is a random variable valued in $L^2(\PP_X)$. We also assume that $X$, $\mathcal{E}$ and $F$ are independent, and that $\mathcal{E} \simdist \mathcal{N}(0, \sigma_{\mathcal{E}}^2)$. We define $Y := F(X) + {\mathcal{E}}$. 
\par \noindent If we assume that our prior distribution for $F$ is proportional to $\exp\left(- \frac{\psi(F)}{2 \sigma_F^2}\right)$ for some functional $\psi: L^2(\PP_X) \to \R_+$, using Bayes' lemma and the independence of $X$, $\mathcal{E}$ and $F$, we obtain the following relation for the likelihood $L(F|(X,Y))$ of $F$ knowing $X$ and $Y$
$$
L(F | (X, Y)) \propto L(F) L( (X,Y) | F ) = L(F) L( (X, \mathcal{E}) | F ) \propto L(F) L(\mathcal{E}) \propto \exp\left(-\frac{1}{2 \sigma_{\mathcal{E}}^2} \left(Y-F(X) \right)^2 - \frac{\psi(F)}{2 \sigma_F^2}\right),
$$
\par \noindent The functional $\psi$ is usually a measure of irregularity such as $\psi(f) := C \int_\R \left(f^{(p)}(x)\right)^2 dx$ for a nonnegative integer $p$. If $(X_i, Y_i)_{1 \leq i \leq N}$ are $N$ independent copies of $(X,Y)$, the likelihood of $F \in L^2(\PP_X)$ given this sample satisfies
$$
L(F | (X_i, Y_i)_{1 \leq i \leq N}) \propto \exp\left(-\frac{1}{2 \sigma_{\mathcal{E}}^2} \sum\limits_{i = 1}^N \left(Y_i-F(X_i) \right)^2 - \frac{\psi(F)}{2 \sigma_F^2}\right). 
$$
Hence, maximizing the likelihood of $f$ amounts to solving the minimization problem
\begin{equation}\label{eq:first_tikho}
\min\limits_{f \in \sspan\limits_{i \in I} (f_i)} \underbrace{\frac{1}{N} \sum\limits_{i=1}^N (y_i - f(x_i))^2}_{\textnormal{estimator of} \ \E\left[(Y-f(X))^2\right]} + \underbrace{\frac{\sigma_{\mathcal{E}}^2}{\sigma_F^2} \frac{1}{N}}_{\textnormal{Tikhonov factor}} \psi(f).
\end{equation}
\par \noindent This shows us that the Tikhonov factor should be proportional to $\frac{1}{N}$ where $N$ is the sample size, which is consistent with the intuition that the larger the sample is, the less we need to regularize to avoid over-fitting.

\subsubsection*{The quadratic case}
\par \noindent In the case where the functional $\psi$ is such that $\psi\left(\sum\limits_{i \in I} w_i f_i \right)$ is a quadratic form in the loadings $(w_i)_{i \in I}$, the minimization problem \eqref{eq:first_tikho} simply amounts to the minimization of a quadratic form. It is the case, for example when $\psi(f) = \int_\R (f^{(p)}(x) - g(x))^2 dx$ for some $g \in L^2(\R)$. We solve the Tikhonov-regularized regression problem by solving the corresponding set of normal equations. 
\par Let $V$ be the matrix defined by $V_{ij} := \frac{1}{N} \sum\limits_{l = 0}^N f_i(x_l) f_j(x_l)$, $i,j \in I$ and $c$ be the vector defined by $c_i := \frac{1}{N} \sum\limits_{l = 0}^N f_i(x_l) y_l$, $i \in I$. We assume that the quadratic form $\psi$ is defined by $\psi(w) := \frac{1}{2} w P w + q w$, $w \in \R^I$. After some algebra, the minimization problem \eqref{eq:first_tikho} amounts to
\begin{equation}\label{eq:second_tikho}
\min\limits_{w \in \R^I} \frac{1}{2} w V w - c w + \lambda \left(\frac{1}{2} w P w + q w \right).
\end{equation}
\par \noindent We obtain the following system of normal equations by differentiating \eqref{eq:second_tikho}
\begin{equation}
(V + \lambda P) w + (\lambda q - c) = 0.
\end{equation}
\subsubsection*{Quadratic forms of interest and measure of smoothness}
\par In the case where the basis functions $(f_i)_{i \in I}$ are B-spline basis functions, measures of smoothness of the form $\psi(f) = \int_\R \left(f^{(p)}\right)^2(x) dx$ for some $p$ can be explicitly derived in terms of the loadings $w$. 
\par To begin with, if $f = \sum\limits_{j = 0}^{k + n} w_j b^{\Gamma}_{j, n}$ has non-zero weights on basis functions that have unbounded support and of extrapolating order higher or equal to $p$, we get $\psi(f) = +\infty$. Therefore, the basis truncation order $t$ should always satisfy $t < p$. In other words, $w_i = 0$ for $0 \leq j \leq n - p$ and $k + p \leq j \leq k + n + 1$. For example, with a penalization order $p=2$, the maximum extrapolation order $t$ should be strictly lower than $2$. We obtain
$$
\int_\R \left(\sum\limits_{j = 0}^{n + k} w_j \left(b^{\Gamma}_{j, n}\right)^{(p)}(x) \right)^2 dx = \sum\limits_{i = 0}^{n + k} \sum\limits_{j = 0}^{n + k} w_i w_j \int_\R \left(b^{\Gamma}_{i, n}\right)^{(p)} (x) \left(b^{\Gamma}_{j, n}\right)^{(p)} (x) dx.
$$
Using the explicit decomposition of $\left(b^{\Gamma}_{j, n}\right)^{(p)}$ onto the B-spline basis of order $n-p$, $\left(b^{\Gamma}_{j, n-p}\right)_{0 \leq i < k + n - p + 1}$, the coefficients of the quadratic form depend on inner products of basis functions of order $n-p$, $P_{ij}:= C \int_\R b^{\Gamma}_{i, n-p} (x) b^{\Gamma}_{j, n-p} (x) dx$, which can be computed exactly using Gauss-Legendre quadrature or any of the other methods to compute inner products of B-splines presented in \cite{VermeulentBartelsHeppler}.
\begin{remark}[Penalization of order $n+1$]
\par We are restricted to a penalization order satisfying $p \leq n$. Using the Dirac comb $\left(b^\Gamma_{j,-1}\right)_{0 \leq j < k}$ introduced in Proposition \ref{prop:b_spline_derivatives_higher_order}, we can penalize the derivative of order $n+1$ in the same fashion. Regarding the inner product of B-splines of order $-1$, we use the convention
\begin{itemize}
\item $\int_\R (b^\Gamma_{j,-1} (x))^2 dx := \frac{1}{2}\left(\frac{1}{\gamma_j - \gamma_{j - 1}} + \frac{1}{\gamma_{j+1} - \gamma_j}\right)$, $1 \leq j < k-1$, 
\item $\int_\R (b^\Gamma_{0,-1} (x))^2 dx := \frac{1}{2}\frac{1}{\gamma_1 - \gamma_{0}}$ \ \ and \ \ $\int_\R (b^\Gamma_{k-1,-1} (x))^2 dx := \frac{1}{2}\frac{1}{\gamma_{k-1} - \gamma_{k-2}}$,
\end{itemize}
which corresponds to the trapezoidal rule. 
\end{remark}
\begin{remark}[Invariance by re-scaling]
A desirable property is that if the sample $(x_j, y_j)_{1 \leq j \leq N}$ and the knots $\Gamma$ are simultaneously affinely transformed, the result of the penalized regression remains the same. 
\par On the one hand, an affine transformation of the $y$-axis affects the regression error term and the penalization term in the same fashion and will not change the shape of the penalized regression. In the other hand, an affine transformation of the $x$-axis only affects the smoothness penalization term and thus its relative importance w.r.t. the regression error. In general, if $g(x / \lambda ) = f(x)$ for some $\lambda >0$, then $g^{(p)}(x/\lambda)/\lambda^p = f^{(p)}(x)$ and $\int_\R \left(f^{(p)}(x)\right)^2 dx = \frac{1}{\lambda^{2 p - 1}} \int_\R \left(g^{(p)}(u)\right)^2 du$. Therefore, for a penalization order $p>0$, we recommend a penalization factor proportional to $\sigma_X^{2p-1}$ where the quantity $\sigma_X$ scales proportionally with $X$, like the mean absolute deviation or the standard deviation. Finally the Tikhonov regularization factor should be of the form
\begin{equation}\label{eq:final_penalization_factor}
K \frac{\sigma_X^{2p-1}}{N},
\end{equation}
where $K$ is independent of $\sigma_X$ and the sample size $N$. 
\end{remark}
\subsubsection*{Numerical experiments with penalized regression}
\par
\begin{figure}[!ht]
	\begin{minipage}[c]{0.45\linewidth}
	\includegraphics[height=5.4cm]{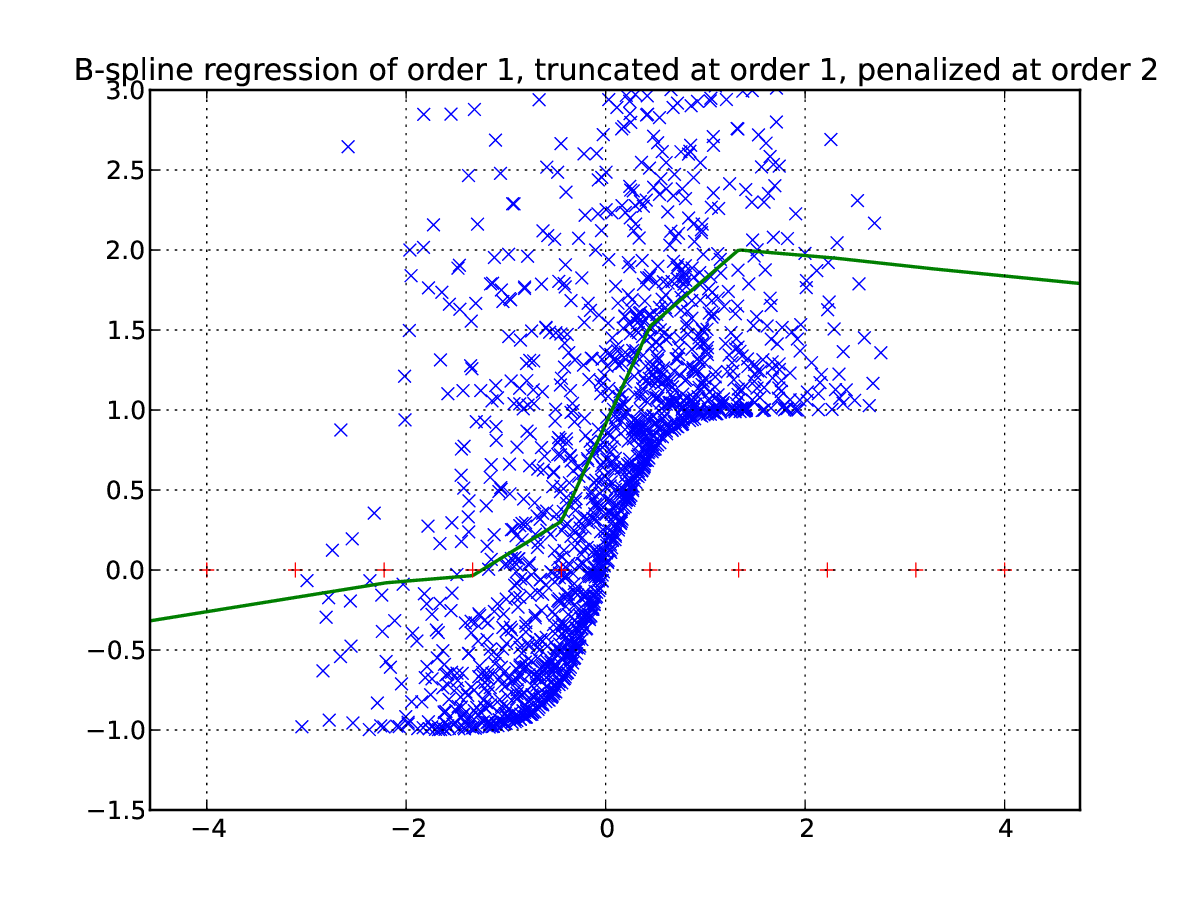}
	\end{minipage} \hfill
	\begin{minipage}[c]{0.45\linewidth}
	\includegraphics[height=5.4cm]{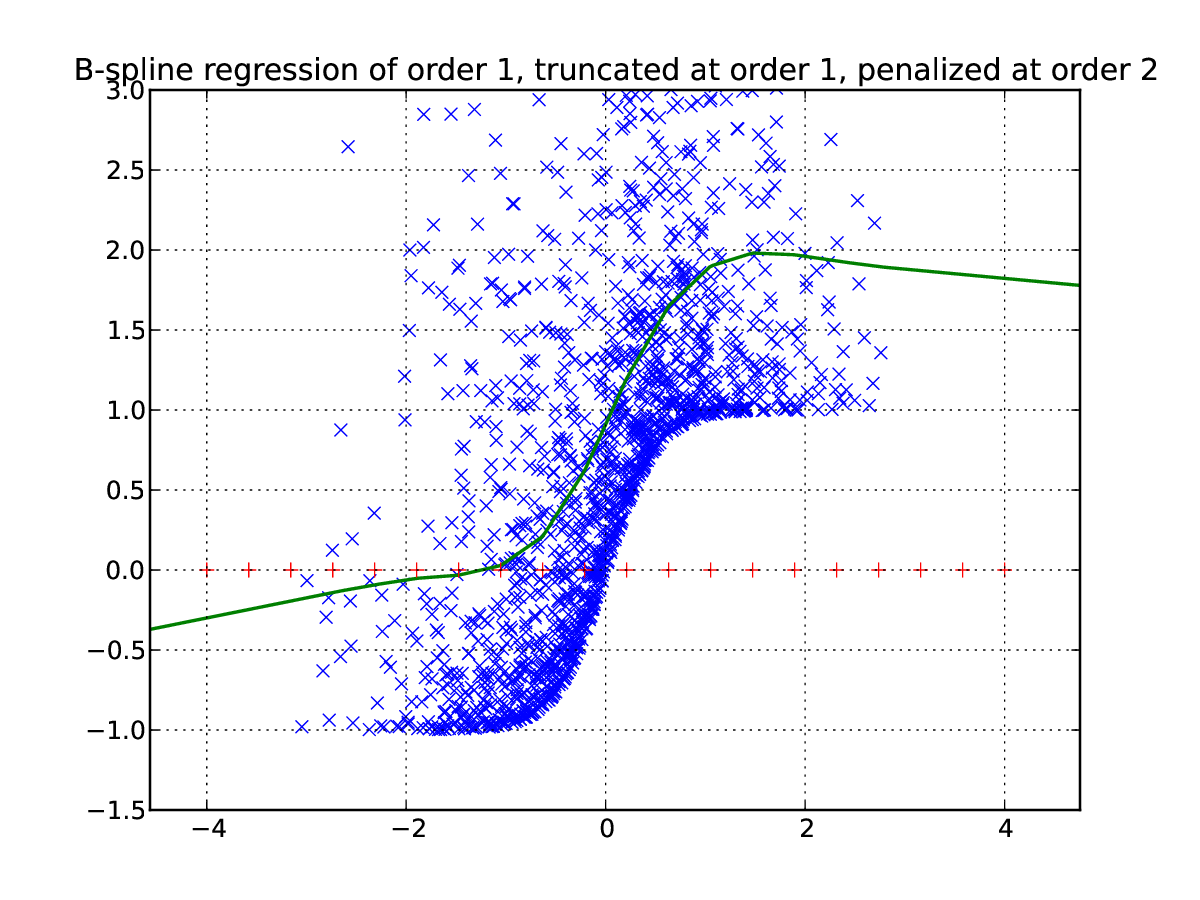}
	\end{minipage}\\
	\begin{minipage}[c]{0.45\linewidth}
	\includegraphics[height=5.4cm]{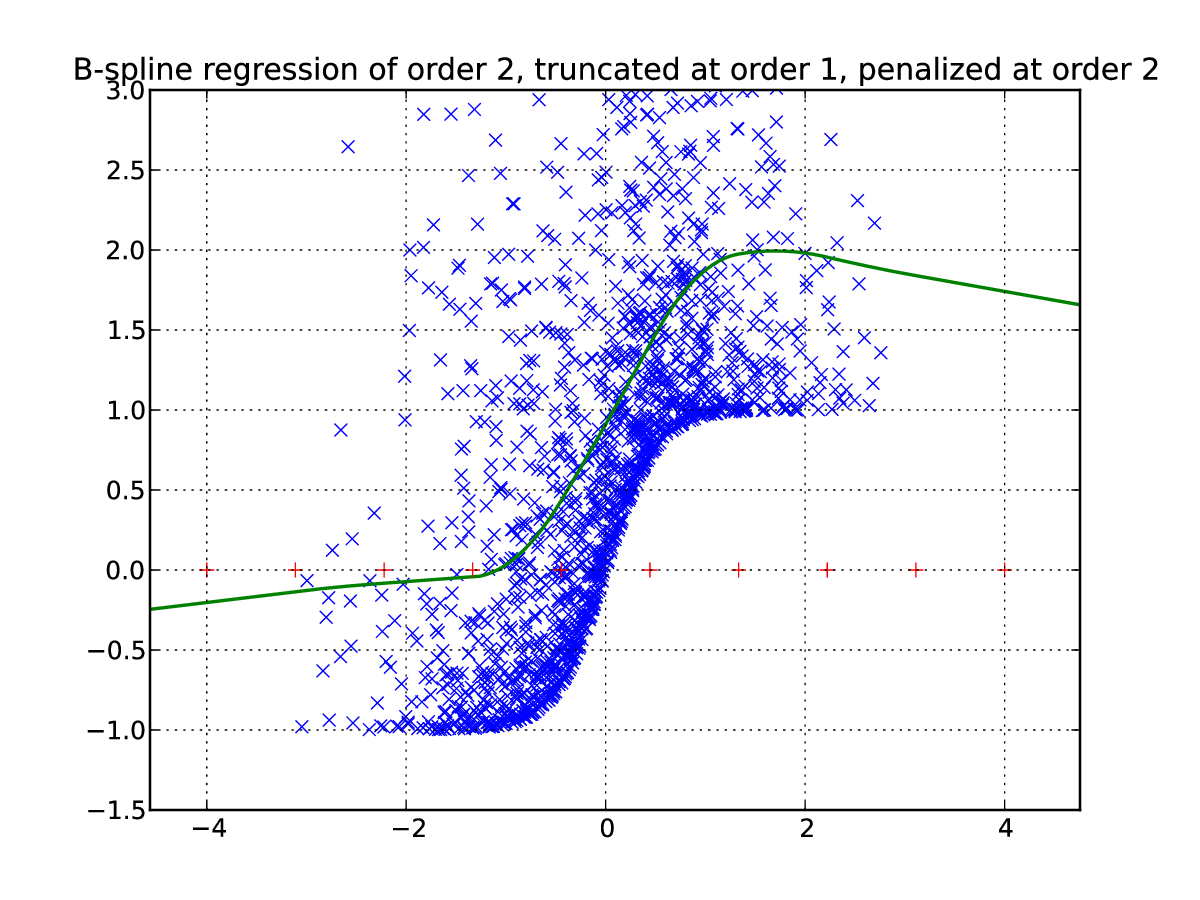}
	\end{minipage} \hfill
	\begin{minipage}[c]{0.45\linewidth}
	\includegraphics[height=5.4cm]{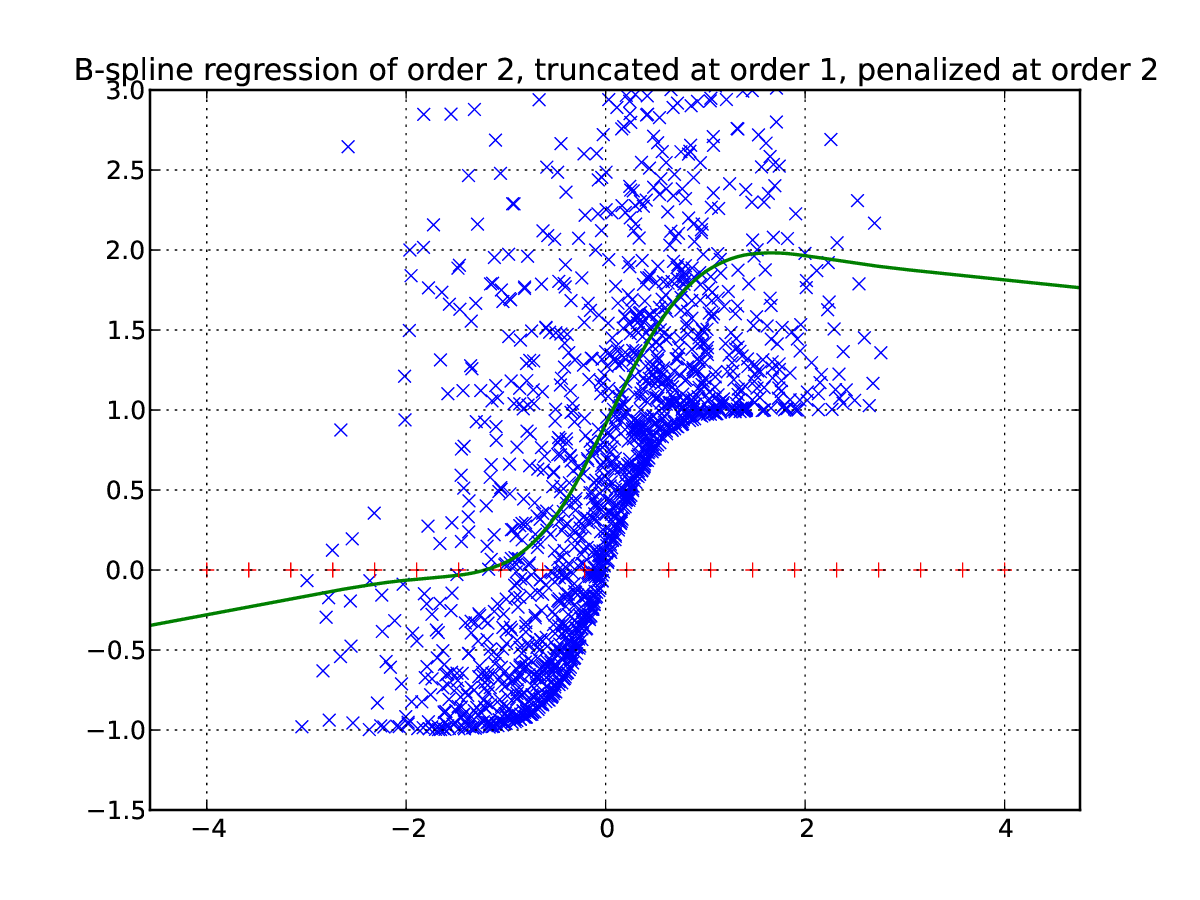}
	\end{minipage}\\
	\begin{minipage}[c]{0.45\linewidth}
	\includegraphics[height=5.4cm]{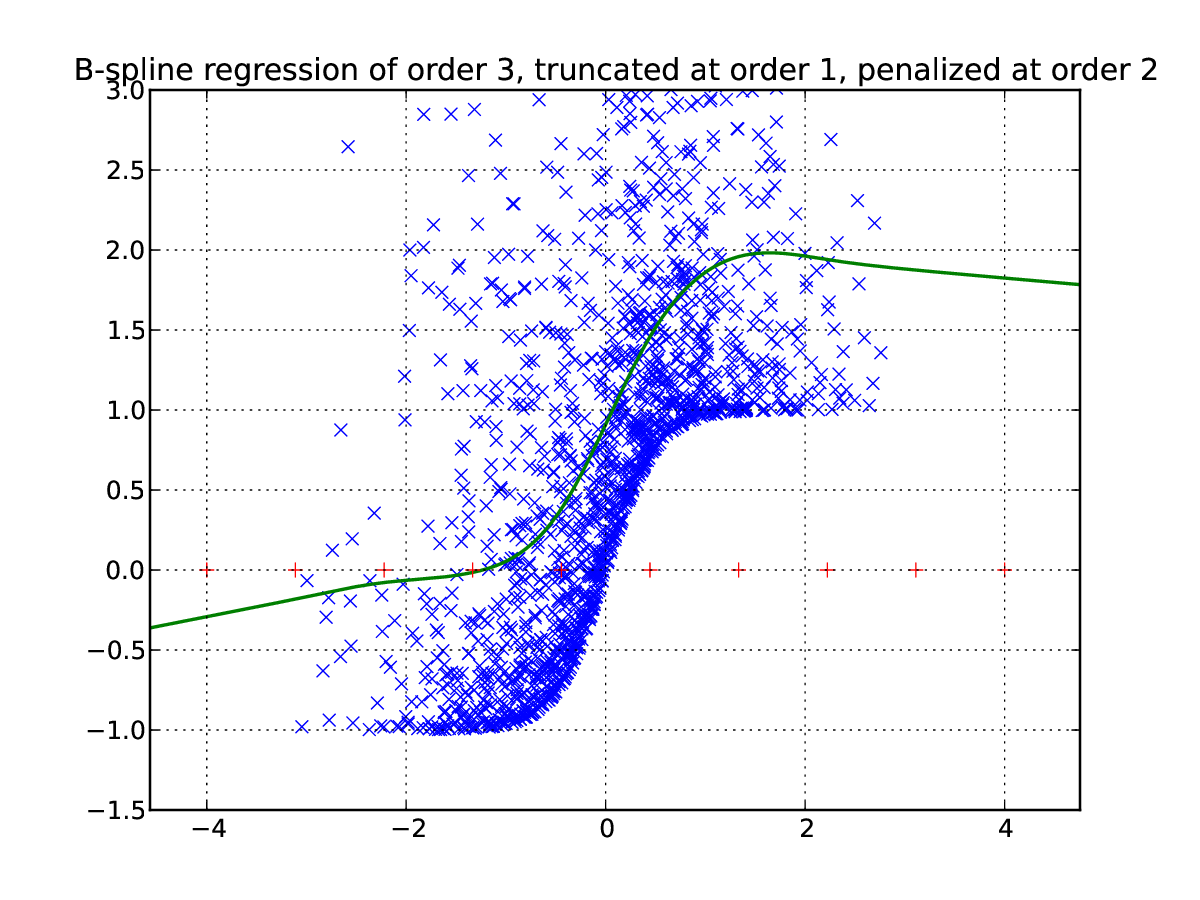}
	\end{minipage} \hfill
	\begin{minipage}[c]{0.45\linewidth}
	\includegraphics[height=5.4cm]{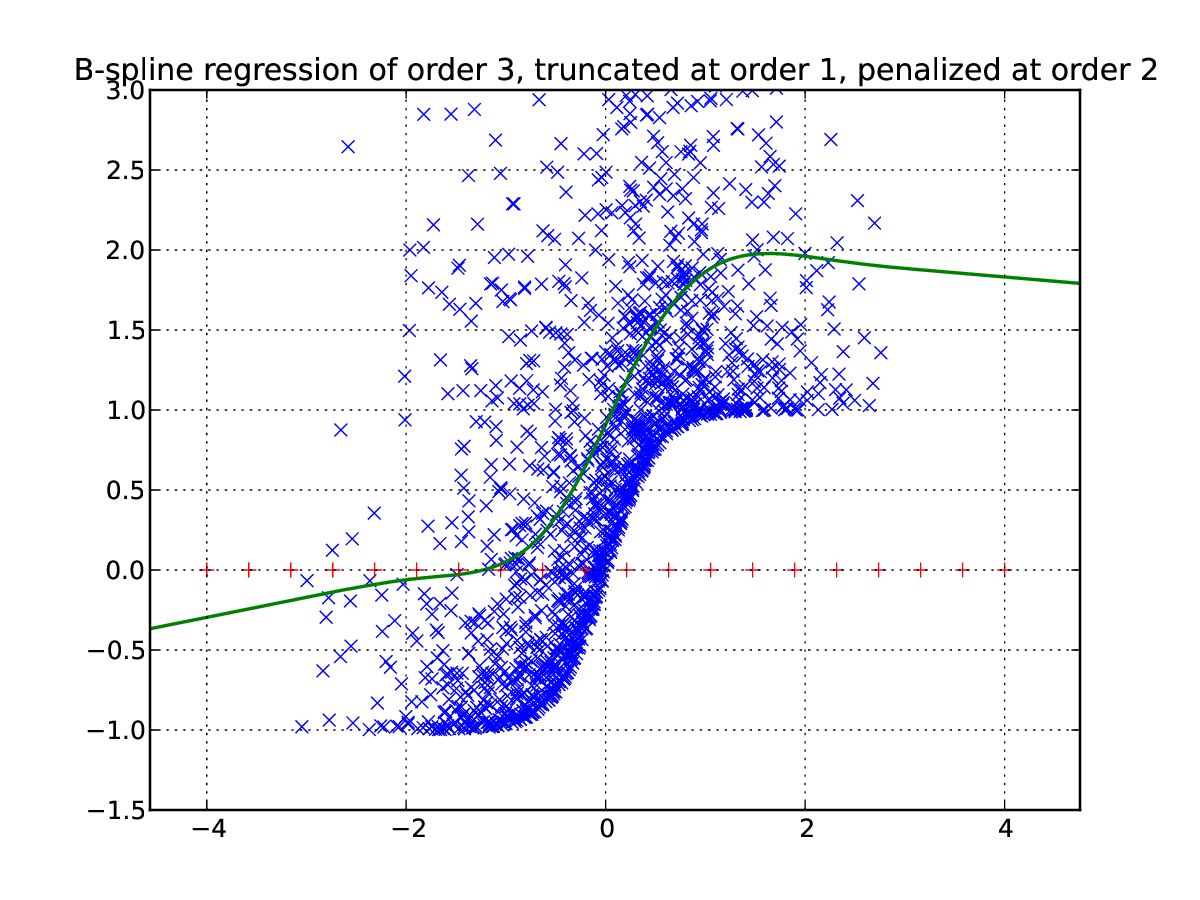}
	\end{minipage}
	\caption{Penalized B-spline regression of a scatter plot sampled from Distribution \eqref{eq:sample_description}. In every case, the penalization order is $p = 2$ and the truncation order is $t = 1$. The Tikhonov penalization factor is $\frac{\sigma_X^{2p - 1}}{N}$ where $N=1600$ is the sample size and $\sigma_X$ the standard deviation of the $X$ sample. We notice that the result does not depend significantly on the spline order. When using a finer discretization grid (right column), the results obtained with different spline orders get closer to each other.}
	\label{fig:regression_tests}
\end{figure}

 In Figure \ref{fig:regression_tests}, we present the penalized regression of the same sample of $(X,Y)$ with B-splines of various orders, various numbers of knots and penalization order $p=2$. In every case, we used a Tikhonov regularization factor of $\frac{\sigma_X^{2p-1}}{N}$. We observe that the results are not very dependent on the spline order, or the number of knots once it is large enough. No additional tuning has been done. For these experiments, the random variables $X$ and $Y$ are defined by
\begin{equation}\label{eq:sample_description}
\begin{array}{ll}
X \simdist \mathcal{N}\left(0, \sigma_X^2 \right)\\
Y := \tanh(2X / \sigma_X) + Z^2, \qquad \textnormal{where} \quad Z \simdist \mathcal{N}(0, \sigma_Z^2) \ \textnormal{ is independent of} \ X, 
\end{array}
\end{equation}
with $\sigma_X = \sigma_Z = 1$. This test case is nonlinear and presents changes of convexity. 
\par
\begin{figure}[!ht]
	\begin{minipage}[c]{0.45\linewidth}
	\includegraphics[height=5.4cm]{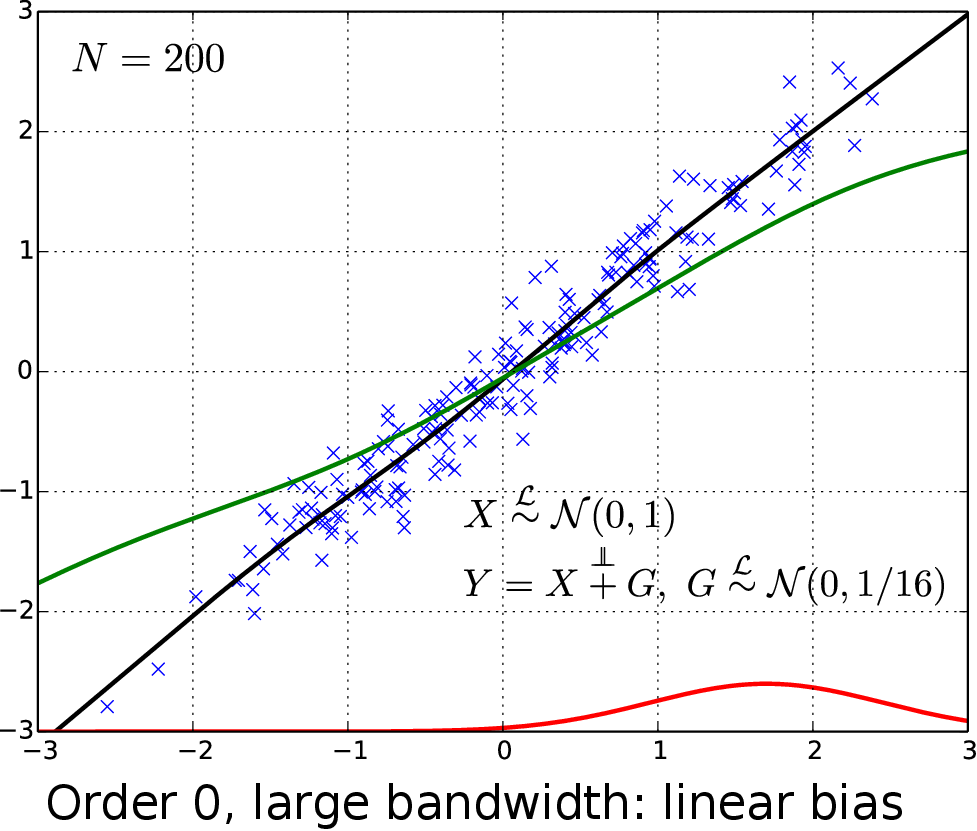}
	\end{minipage} \hfill
	\begin{minipage}[c]{0.45\linewidth}
	\includegraphics[height=5.4cm]{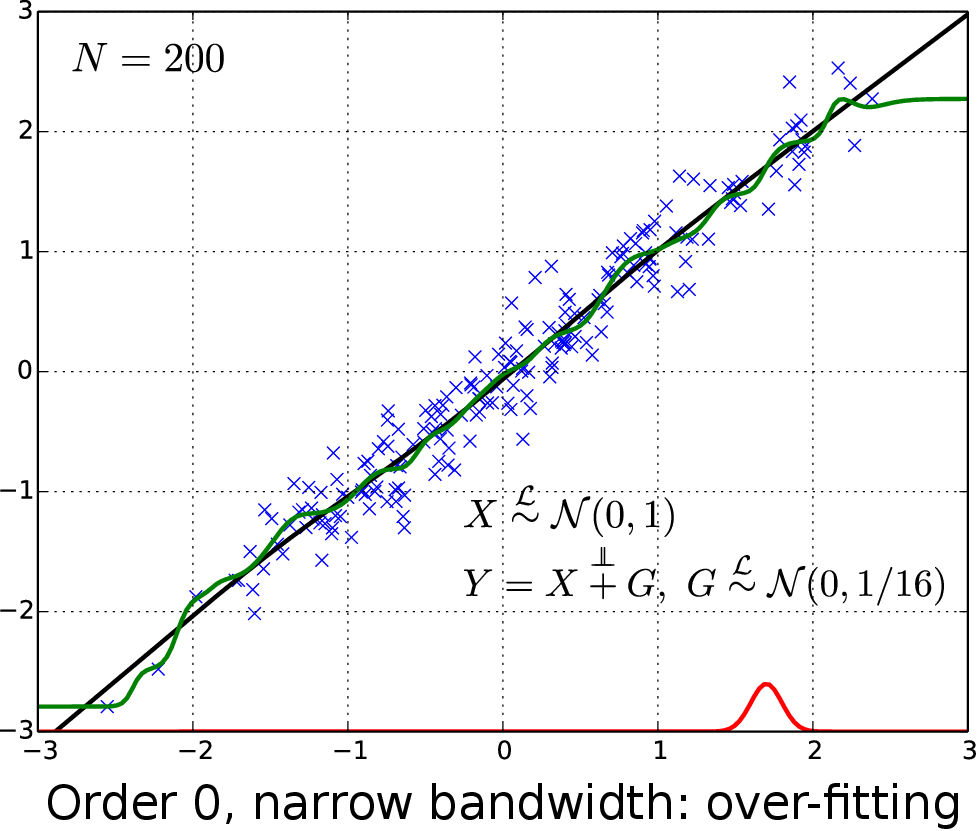}
	\end{minipage}\\
	\begin{minipage}[c]{0.45\linewidth}
	\includegraphics[height=5.4cm]{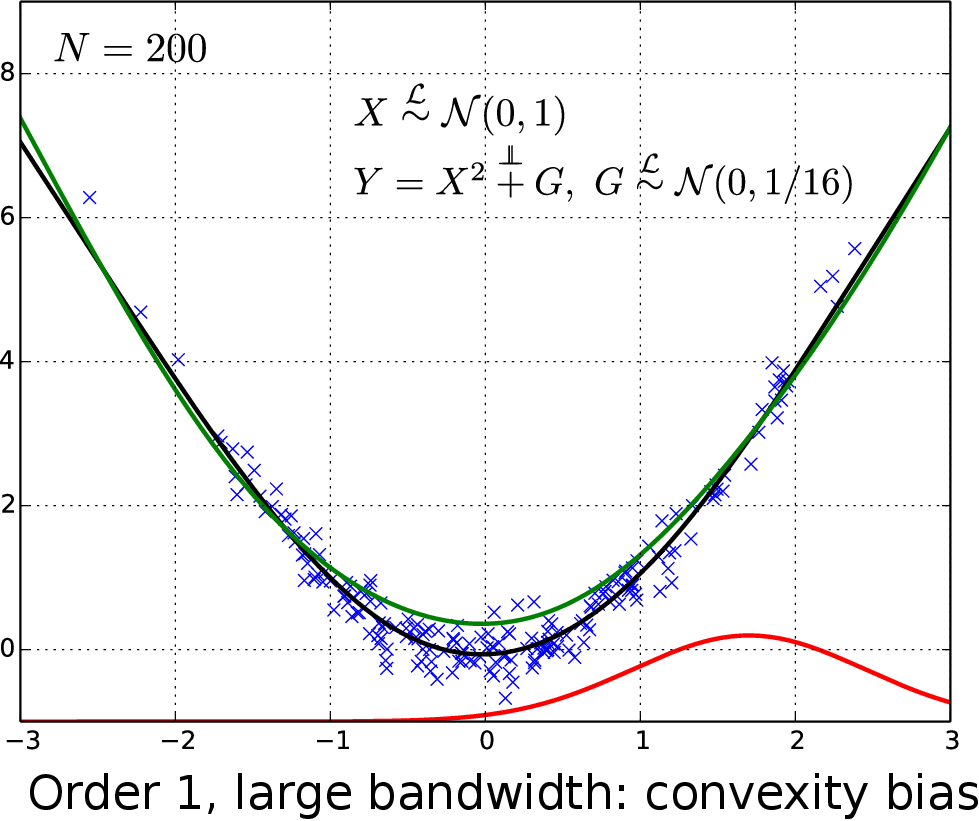}
	\end{minipage} \hfill
	\begin{minipage}[c]{0.45\linewidth}
	\includegraphics[height=5.4cm]{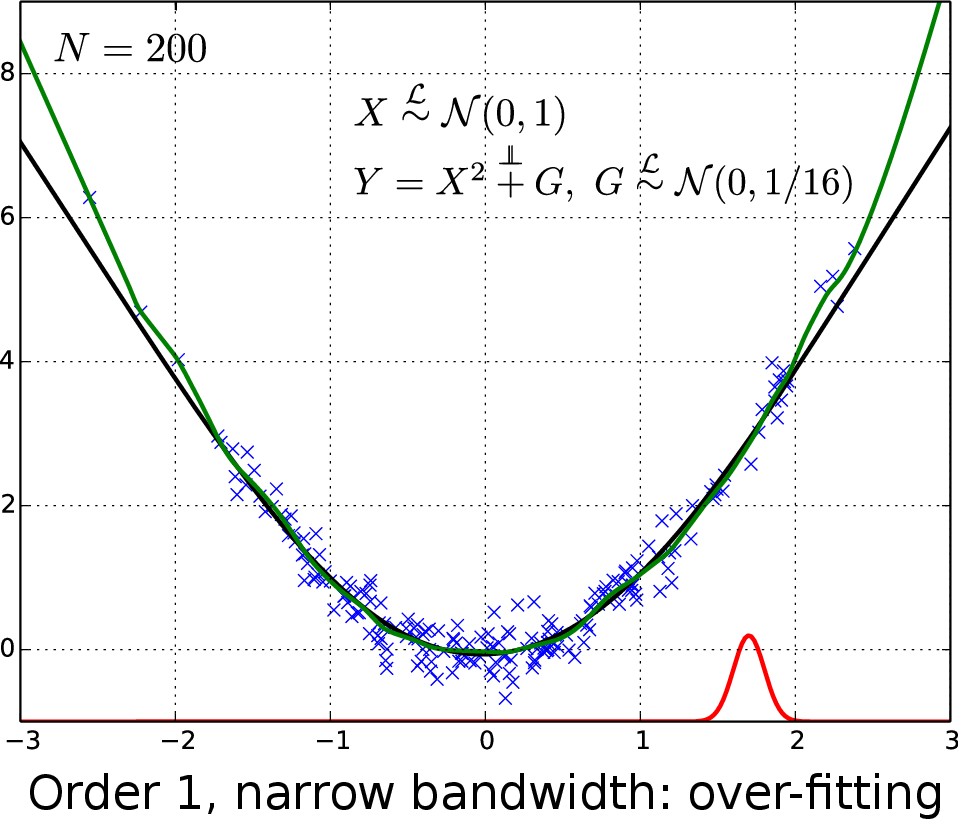}
	\end{minipage}\\
	\begin{minipage}[c]{0.45\linewidth}
	\includegraphics[height=5.4cm]{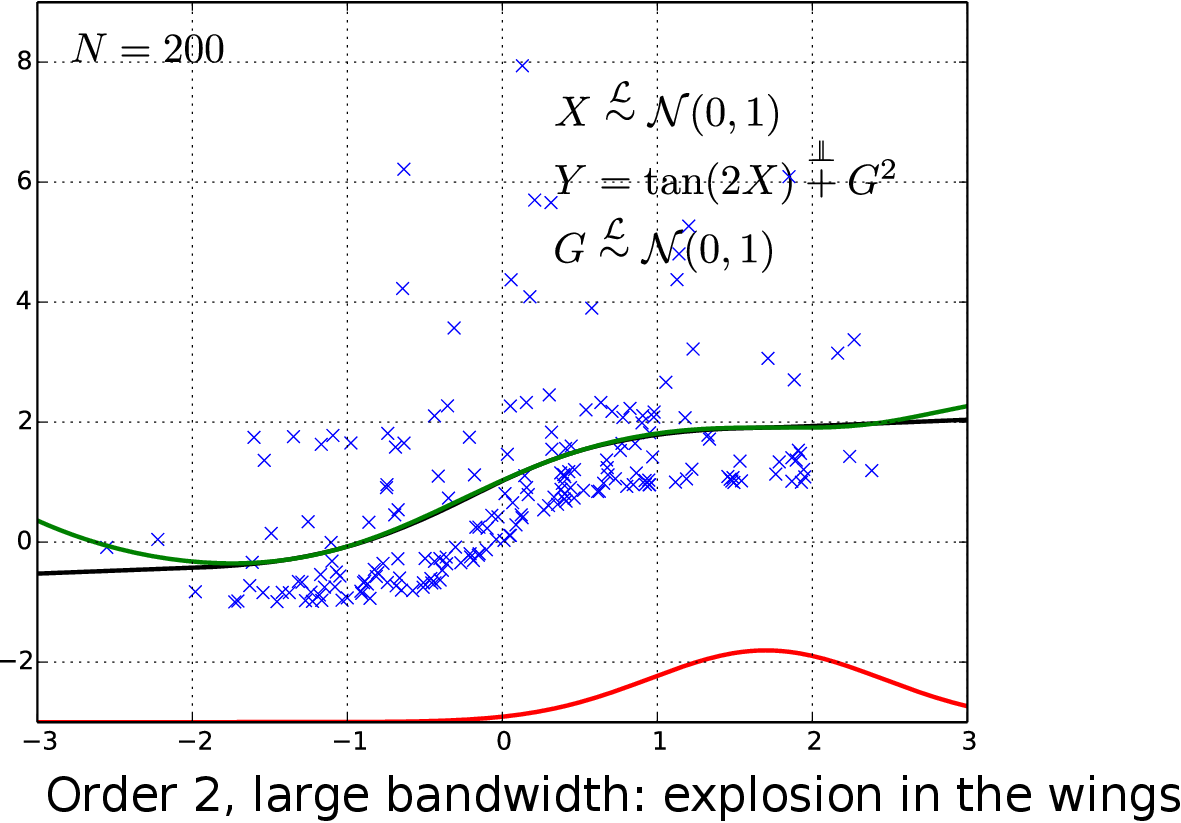}
	\end{minipage} \hfill
	\begin{minipage}[c]{0.45\linewidth}
	\includegraphics[height=5.4cm]{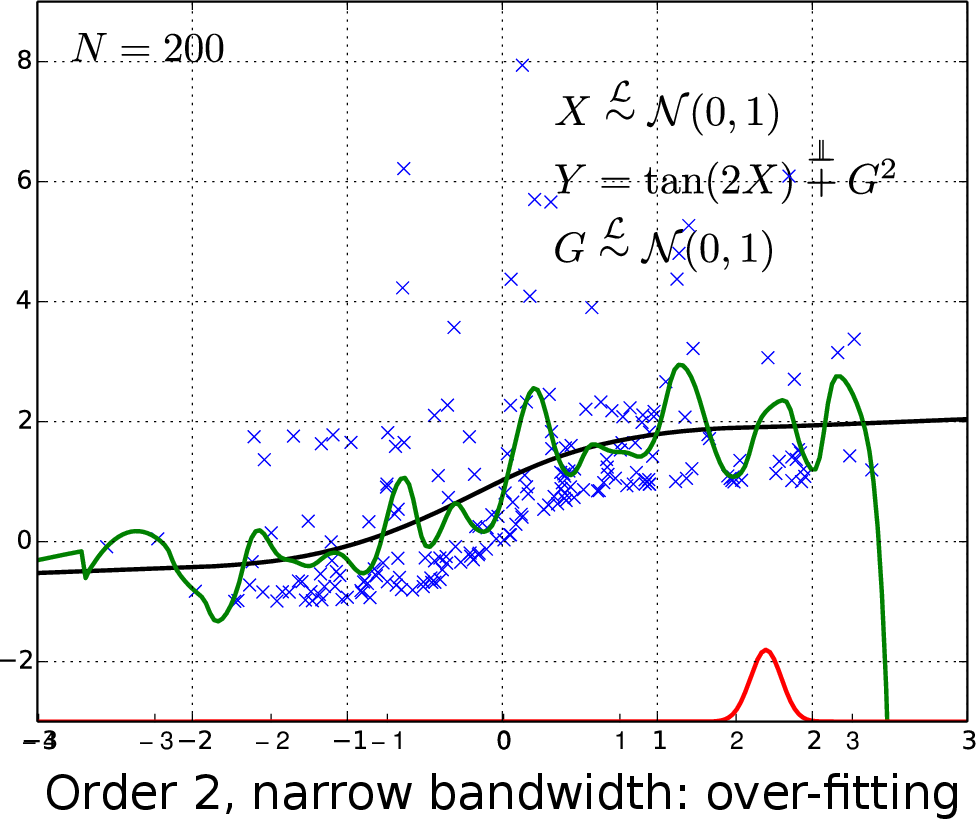}
	\end{minipage}
	\caption{Illustration of the different flaws on non-parametric regression. In each figure the black line corresponds to a second-order B-spline regression with $10$ knots evenly spaced between $-2.5$ and $2.5$, with a Tikhonov-regularization of order $p=2$ and a regularization factor of $\frac{\sigma_X^{2p-1}}{N}$. The green line is the result of the non-parametric regression. The first row corresponds to order-$0$ non-parametric regression (local mean); the second row corresponds to order-$1$ non-parametric regression (local linear regression); and the last row corresponds to order-$2$ non-parametric regression (local second-order polynomial fit). }
	\label{fig:non_parametric_regression_tests}
\end{figure}

 In Figure \ref{fig:non_parametric_regression_tests}, we illustrate the principal flaws of non-parametric regression of different orders. We display the shape of the smoothing kernel used for the regression. Order-$0$ local regression (local mean) will suffer from a general bias of order $1$ in the case of a large bandwidth. Namely, when the bandwidth tends to infinity, the estimated value for $\E[Y | X=x]$ converges to $\E[X]$ regardless of $x$. In the case of too narrow bandwidth, order-$0$ local regression over-fit the data (See the first row in Figure \ref{fig:non_parametric_regression_tests}). Similarly, order-$1$ local regression (local linear regression) suffers from an order-$2$ bias in the case of a large bandwidth and over-fits in the case of a narrow bandwidth (second row in Figure \ref{fig:non_parametric_regression_tests}). Finally, local regression of order $2$ will generally present exploding results on the wings. 
\par Penalized B-spline regression has proven to be a rather robust regression method in comparison with classical non-parametric approaches. Non-parametric regression methods are very sensitive to parameters such as the regression order, the selection of the bandwidth, the shape of the smoothing kernel, and give poor control on extrapolation. Moreover, the piecewise polynomial form of B-splines allows for a natural representation of the regression result in memory and a natural way to evaluate it at new values. It has a solid theoretical foundation as a maximum likelihood estimator of the conditional expectation. More importantly, we will see that unlike non-parametric regression, B-splines allow to account for linear shape constraints such as non-negativity, monotonicity and convexity, and linear integral constraints. Such linear constraints come at practically no cost as the regression then amounts to a quadratic program. 
\subsection{Compatibility with the marginal distributions}
\par A common application of multiple regression is the estimation of the conditional expectation of a random variable $Y$ given another random variable $X$ from a scatter plot of the joint distribution.
\par However, it often happens that additional information is available. For example, it is common that we completely know the marginal distributions of $X$ and $Y$. If $X$ and $Y$ are two real $L^1$ random variables and $f: \R \to \R$ is a measurable function such that $\E[Y|X] = f(X)$ a.s. then the following conditions hold:
\vspace{2mm}
\par \noindent \begin{minipage}[l]{0.60\linewidth}
\begin{subequations}\label{eq:compability}
  \begin{alignat}{1}
    \! &\E[Y] = \int_\R f(x) d\PP_X(x) \label{eq:compability:expectation} \\[2mm]
    \! &f(x) \ \textnormal{is} \ \PP_X \textnormal{-a.s. in the convex hull of} \ \supp(\PP_Y) \label{eq:compability:convexhull} \\[2mm]
    \! &\textnormal{For any nonnegative convex function} \ \phi, \nonumber \\
    \! &\qquad \qquad \qquad \qquad \qquad \quad \E[\phi(f(X))] \leq \E[\phi(Y)]. \label{eq:compability:jensen}
  \end{alignat}
\end{subequations}
\end{minipage}
\hfill
\begin{minipage}[l]{0.33\linewidth}
\includegraphics[width=5cm]{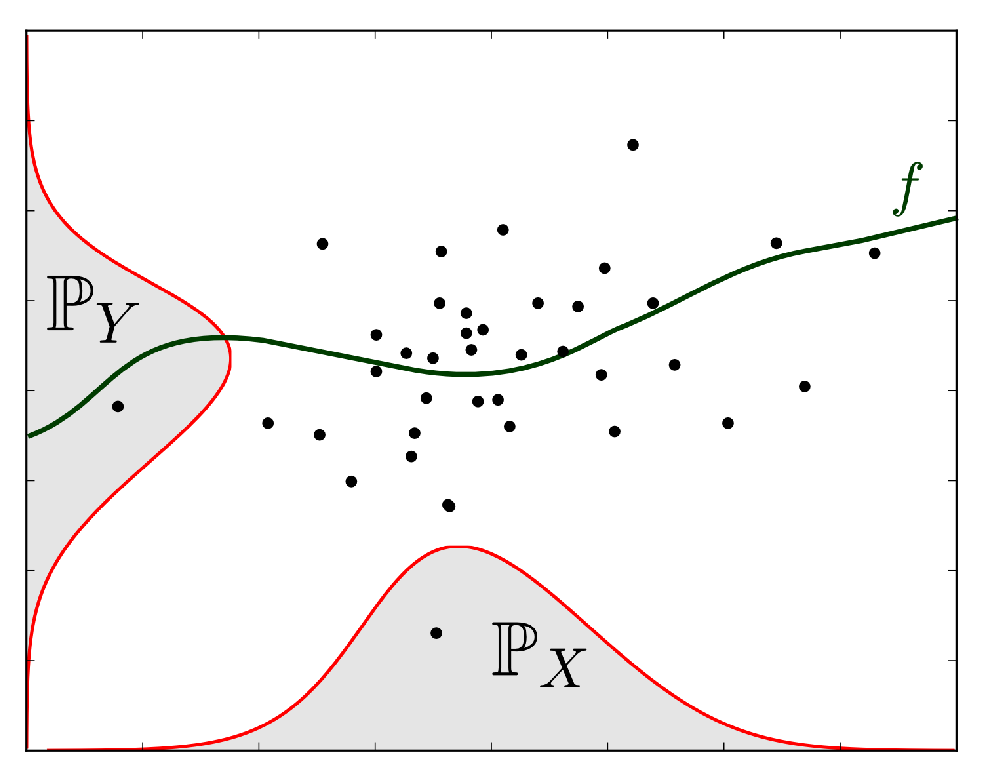}
\end{minipage}
\vspace{2mm}
\par \noindent Therefore, from a Bayesian point of view, it does not make sense to consider an estimate of the conditional expectation that does not satisfy these properties. 
\begin{itemize}
\item The first condition \eqref{eq:compability:expectation} amounts to a linear integral equality constraint.
\item The second one \eqref{eq:compability:convexhull} consists of a set of linear inequality constraints. (In practice, it often amounts to a nonnegativity constraint in the regression.)
\item Condition \eqref{eq:compability:jensen} means that $Y$ dominates $f(X)$ for the convex order. A consequence is that $\E\left[f(X)^2\right] \leq \E\left[Y^2\right]$, which is a quadratic inequality constraint.
\end{itemize}
\par The resulting constrained optimization problem can be formulated as a second-order cone program. Details on this special class of optimization problems is provided in Section \ref{sec:shape_constraints}.
\begin{remark}[Estimation of the conditional median rather than the conditional expectation]
\par In \cite{HeNgConstrainedSplines}, He and Ng proposed a constrained $L^1$ regression technique based on B-splines. More precisely, the quantity of interest that is parameterized with a spline is a conditional quantile distribution $\PP\left[Y \leq g_\tau(x)\middle| X = x \right]$, and $\tau = 1/2$ corresponds to the conditional median. 
\end{remark}
\subsection{Shape constraints and second-order cone programming}\label{sec:shape_constraints}
\par In this section, we first give some background on second-order cone programming and quadratic programming. Then, we review the shape constraints on B-splines that qualify as second-order cone constraints. 
\subsubsection*{Second-order cone programming}
\par \noindent A second-order cone program is a minimization problem of the form
\begin{equation}\label{eq:socp}
\begin{array}{ll}
\textnormal{minimize}	& f^T x\\
\textnormal{subject to} & \left\| A_i x + b_i \right\|_2 \leq c_i^T x + d_i, \quad i =1, \cdots, N,
\end{array}
\end{equation}
where $A_i \in M_{n_i-1,n}(\R)$, $b_i \in \R^{n_i-1}$, $c_i \in \R^n$ and $d_i \in \R$. Second-order cone constraints, of the form $\left\| A_i x + b_i \right\|_2 \leq c_i^T x + d_i$ reduce to
\begin{itemize}
\item linear inequality constraints if $n_i = 1$, ($0 \leq c_i^T x + d_i$),
\item quadratic constraints if $c_i = 0$, ($\| A_i x + b_i \|_2^2 \leq d_i^2$). 
\end{itemize}
\par \noindent Moreover, in the case where the objective function itself is a positive definite quadratic form, we can recast it as a second-order cone program by appending an additional scalar $t$ to the optimization variable.
\par \noindent The optimization problem
$$
\begin{array}{rl}
\textnormal{minimize}	& x^T P^T P x + 2 q_0^T x\\
\textnormal{subject to} & \textnormal{a collection of second-order cone constraints}
\end{array}
$$
where $P$ is an invertible matrix, amounts to the minimization problem
$$
\begin{array}{rl}
\textnormal{minimize}	& t\\
\textnormal{subject to} & \textnormal{the same collection of second-order cone constraints on} \ x\\
\textnormal{and}		& \left\| P x + P^{-1} q_0 \right\|_2 \leq t
\end{array}
$$
where the new optimization variable is $\left(x_1, \cdots, x_n, t\right) \in \R^{n+1}$. Highly efficient software packages to solve second-order cone program are available, such as CVXOPT \cite{cvxopt}, Mosek \cite{mosek} or CPLEX \cite{ilogcplex}.
\subsubsection*{Shape constraints on B-splines}
\par As we have seen in Section \ref{sec:spline_section}, if $k$ is a nonnegative integer, $\Gamma := \left\{ \gamma_0 \leq \gamma_1 \leq \cdots \leq \gamma_{k-1} \right\}$ is a sorted collection of $k$ knots and if $n \leq k$, the B-spline basis functions of order $n$ are nonnegative functions. Hence, the nonnegativity of each one of the the loadings is a sufficient condition for nonnegativity. It is also a finite set of linear constraints. This condition happens to be necessary for B-splines of order $0$ and $1$ as well as for the Dirac comb $\left(b^\Gamma_{i, -1}\right)_{0 \leq i < k}$ introduced in Proposition \ref{prop:b_spline_derivatives_higher_order}. 
\par As we have seen in Proposition \ref{prop:b_spline_derivatives}, derivatives of B-spline basis functions are explicitly decomposed onto the basis of lower order. Thus nonnegativity constraints on the first and second derivatives of B-splines translate into monotonicity and convexity constraints. 
\begin{remark}
There is no simple sufficient and necessary condition for spline nonnegativity of order $n \geq 2$. However, in \cite{PappAlizadeh} Papp and Alizadeh devised a method to handle the global nonnegativity constraints on B-splines without restraining to the case of nonnegative coefficients on a nonnegative basis, while remaining within the scope of second-order cone programs. In this article, we settle for the sufficient condition mentioned above.
\end{remark}
\subsubsection*{Other linear constraints}
\par Equality and inequality constraints on the value of a B-spline or one of its derivatives at a certain point obviously qualify as linear constraints. It is also the case for inequality and equality constraints on limits of a B-spline or its derivatives at $-\infty$ or $+\infty$. 
\par Regarding integral constraints, if $f_w = \sum\limits_{j = 0}^{k + n} w_j b^{\Gamma}_{j, n}$ is a B-spline of order $n$ and $\mu$ a given locally finite measure, we have $I(w) := \int_\R f_w(x) d\mu(x) = \sum\limits_{j = 0}^{k + n} w_j \int_\R b^{\Gamma}_{j, n}(x) d\mu(x)$. Therefore, (if the quantities $\int_\R b^{\Gamma}_{j, n}(x) d\mu(x)$ are known for $0 \leq j < k + n + 1$), equality and inequality constraints on $I(w)$ qualify as linear constraints.
\subsubsection*{Hierarchy of equality constraints: a modified Moore-Penrose pseudoinverse}
\par Shape-constrained B-splines can be used as an interpolation method rather than a multiple regression method. In this case, there is temptation to consider the interpolation condition as firm equality constraints, and to use a measure of smoothness for the objective function in the resulting second-order cone program. However, we can encounter feasibility issues when using this approach. The input data could be unreachable with the given knots and spline order. 
\par A more robust approach is to use all the degrees of freedom to achieve a least-square fit of the data points, and among the solutions of this problem, maximize smoothness. If interpolation is feasible, it will be achieved and the most regular interpolator will be returned. 
\vspace{2mm}
\par The singular value decomposition of a real matrix $A \in M_{l,p}(\R)$ is the decomposition $A = U D V^*$ where $U$ and $V$ are (complex) unit matrices and $D$ is a $l \times p$ diagonal matrix. We denote by $D^+$ the $p \times l$ diagonal matrix obtained by inverting non-zero entries of $D$ and transposing it. It satisfies the following properties:
\begin{itemize}
\item For $b \in \R^l$, $\widehat{x}:= (V D^+ U^*) b$ is a solution to $\min\limits_{x \in \R^p} \| Ax - b\|_2$. 
\item If the minimization problem $\min\limits_{x \in \R^p} \| Ax - b\|_2$ has multiple solutions, $\widehat{x}$ is the one which has the minimal Euclidean norm. $A^+ := V D^+ U^*$ is called the Moore-Penrose pseudoinverse of $A$.
\end{itemize}
\par \noindent One could prefer to minimize another quadratic form $x \mapsto x^T Q x$, different from the Euclidean norm. If $Q$ is positive definite and $Q = G^T G$ is its Cholesky decomposition, we define $\widehat{x}:= G^{-1}\left(AG^{-1}\right)^+ b$. Using the properties of the Moore-Penrose pseudoinverse mentioned above, we find
\begin{itemize}
\item $\widehat{u} := G \widehat{x}$ minimizes $\left\|A G^{-1}u - b \right\|_2$, which implies that $\widehat{x}$ minimizes$ \|Ax - b \|_2$,
\item $Gx$ has a minimal Euclidean norm, and thus $x^T Q x$ is minimal. 
\end{itemize}
The matrix $A_Q^+:= G^{-1} \left(AG^{-1}\right)^+$ is the pseudoinverse of $A$ that minimizes the quadratic form $Q = G^T G$. In general, we would always recommend to use this approach when all constraints are linear equality constraints, in order to give a best fit result in the case of infeasibility. However, the same analysis cannot be carried out in presence of inequality constraints. We refer to \cite{hierarchQuadraticProgramming} for a thorough review of methods to handle hierarchies of constraints with more general quadratic programs.
\section{Application to Guyon and Henry-Labordère's particle method}\label{sec:particle_method}
\par In this section, we take on the application of the shape-constrained regularized B-spline regression to the calibration of the leverage function in stochastic local volatility models.
\par Let $S_t$ be the price of a risky asset at time $t$. We assume for the sake of simplicity that the asset does not pay any dividend and that interest and repo rates are zero. Then, arbitrage pricing theory tells us that under any risk-neutral probability, $S_t$ is a martingale. 
\par Knowledge of the call and put option prices of all strikes and maturities is equivalent to the knowledge of the risk-neutral densities of $S_t$ for every maturity $t$. The celebrated Local Volatility Model \cite{DupirePricingSmile} is the only Markov diffusion to match the corresponding continuum of marginal distributions. The local volatility function is given by Dupire's stripping formula. (We use the Bachelier convention for instantaneous volatilities, that is, $dS = \sigma dW$, rather than the lognormal convention $dS = S \sigma dW$.)
\begin{equation}\label{eq:stripping_form}
\sigma_{\textnormal{Dup}}^2(T, x) = \frac{\frac{\partial C}{\partial T}}{\frac{1}{2} \frac{\partial^2 C}{\partial K^2}}.
\end{equation}
However, it is an arbitrary choice for the modeling of transition probabilities. It may not be a good model to price and hedge products that depend on these transition probabilities (see \cite{DupireUTV}). Pure stochastic volatility models, such as SABR \cite{SABRSmallMaturity}, the Heston model \cite{HestonModel} or Bergomi's model \cite{SmileDynamics2} are a first attempt of the modeling of these transition probabilities. A potential problem is that they do not have enough degrees of freedom to match all quoted vanilla option prices. A widespread approach \cite{RenMadanQianQian} is the embedding of an additional level-dependent function $l(t,x)$, the leverage function, into the diffusion equation: 
\begin{equation}\label{eq:slv_sde}
dS_t = a_t l(t, S_t) dW_t.
\end{equation}
The local volatility term $l(t,S_t)$ is not only used as a means to achieve exact calibration to the market prices. It also allows one to recover the implied volatility skew using a combination of the contribution of the spot-vol correlation and the contribution of the local volatility term.
\subsection{Calibration of the leverage function}
Let us now consider the situation in which the Dupire local volatility is already calibrated. Equivalently, the entire implied volatility surface is known and therefore the risk-neutral marginal distributions $\PP_{S_t}$ are known for any $t \geq 0$. The process $a_t$ is a pure stochastic volatility process, which has also been determined. It only depends on $S_t$ through the correlation of its driving process(es) with the Brownian motion $W_t$. We now tackle the calibration of the leverage function $l(t,S_t)$. 
\par The calibration condition arises from Gyöngy's Markov projection theorem \cite{GyongyMarkovian}, which tells us that the so-defined process $S_t$ has the same marginals as Dupire's Local Volatility Model $d\widetilde{S}_t = \sigma_{\textnormal{Dup}}\left(t,\widetilde{S}_t\right) d\widetilde{W}_t$ if and only if 
\begin{equation}\label{eq:calibration_condition}
\sigma_{\textnormal{Dup}}^2(t, x) = l(t,x)^2 \E\left[ a_t^2 \middle| S_t = x \right]. 
\end{equation}
\par If the leverage function satisfies this condition, then $(S_t, a_t)$ is the solution to a non-linear stochastic integro-differential equation
\begin{equation}\label{eq:nonlinear_sde}
dS_t = \frac{\sigma_{\textnormal{Dup}}\left(t, S_t\right)}{\sqrt{\E\left[a_t^2 \middle| S_t \right]}} a_t dW_t.
\end{equation}
\par \noindent A theoretical study on the existence of solutions to \eqref{eq:nonlinear_sde} was carried out in \cite{AbergelTachet} by Abergel and Tachet from a PDE viewpoint in the case where $a$ is a one-dimensional It\^o process (see Appendix \ref{sec:abergel_tachet}). Regarding its numerical treatment, in Article \cite{GuyonLabordereParticular}, Guyon and Henry-Labordère devised a purely forward Monte Carlo method to integrate \eqref{eq:nonlinear_sde} and calibrate the leverage function $l(t,x)$. Let us also mention the related work of Van der Stoep, Grzelak and Oosterlee \cite{StoepGrzelakOosterlee}. 
\vspace{2mm}
\par Let $T$ be the horizon maturity for the calibration and $t_0 = 0 < \cdots < t_n = T$ a subdivision of $[0, T]$. We assume that the Dupire local volatility function $\sigma_{\textnormal{Dup}}$ is already calibrated and that the model parameters for $a_t$ are already fixed. The calibration procedure proceeds as follows:
\vspace{2mm}
\begin{samepage}
\begin{breakbox}
\par \noindent \textbf{Monte Carlo calibration of the leverage function \cite{GuyonLabordereParticular}}
\par \noindent For every $0 \leq k < n$, simulate $N$ independent draws of $S_{t_{k+1}}, a_{t_{k+1}}$ with a single Euler step (or another stepping scheme) from
\begin{itemize}
\item the $N$ draws of the previous time step of $S_{t_k}, a_{t_k}$
\item the calibrated leverage function $x \mapsto l(t_k,x)$. 
\end{itemize}
We ensure the calibration condition \eqref{eq:calibration_condition} for the next time step by setting
$$
l(t_{k+1}, x) := \frac{\sigma_{\textnormal{Dup}}(t_{k+1}, x)}{\sqrt{\E \left[a_{t_{k+1}}^2 \middle| S_{t_{k+1}} = x \right]}}. 
$$
\end{breakbox}
\end{samepage}
\vspace{2mm}
\par At each time step, the computation of the diffusion coefficient for a given path integrates contributions from all the other simulated paths via the estimation of the conditional expectation of $a_t^2$ knowing $S_t$. In this estimation, in addition to the scatter plot, the risk-neutral probability distribution of $S_t$ is known (from the already interpolated implied volatility surface). Furthermore, the distribution of $a^2_t$ in most stochastic volatility models if also a given. For example, in the case of the Heston model, it is a noncentral chi-squared distribution. In the cases of the SABR model and Bergomi's model, $a^2_t$ is lognormal. Hence, we are in the situation studied in the previous section, where the marginal distributions are known. 
\begin{enumerate}[(a)]
\item Condition \eqref{eq:compability:expectation} results in an equality constraint for the instantaneous forward variance: $\E\left[a_t^2\right] = \int_\R f(x) d\Q_{S_t}(x) = \sum\limits_{i=0}^{k+n}w_i \int_\R b^\Gamma_{i,n}(x) d\Q_{S_t}(x)$.
\item The convex hull condition \eqref{eq:compability:convexhull} amounts to a nonnegativity constraint. 
\item A consequence of the third compatibility condition \eqref{eq:compability:jensen} is the quadratic inequality constraint
\begin{equation}\label{eq:quadratic_constraint_slv}
\E\left[a_t^4\right] \geq \E\left[f(S_t)^2\right] = \sum\limits_{i=0}^{k + n} \sum\limits_{j=0}^{k + n} w_i w_j \int b^\Gamma_{i,n}(x) b^\Gamma_{j,n}(x) d\Q_{S_t}(x).
\end{equation}
\end{enumerate}
\par In Section \ref{sec:particle_method_numerical}, we present numerical experiments with the shape-constrained particle method for the calibration. We show that adding the shape constraints allows us to significantly reduce the number of Monte Carlo runs necessary to reach a desired level of accuracy. 
\begin{remark}[Evaluation and representation of B-splines for the particle method]
\begin{itemize}
\item In the case of the particle method, the resulting B-splines are evaluated at the very same points where the basis functions where evaluated in the first place. Hence, we should keep the corresponding values and simply sum over the already evaluated basis functions rather than relying on backward evaluation schemes.
\item The basis functions are evaluated on a Monte Carlo sample. In this context, it is typically beneficial to pre-compute the piecewise polynomial representation of the basis for the evaluation on this large number of points.
\item Another advantage for the piecewise polynomial representation is that it is the best suited memory representation of the calibrated leverage function for any future use, which does not rely on the definition of the B-splines. 
\end{itemize}
\end{remark}
\par \noindent Implementing the linear equality condition on the instantaneous forward variance $\E\left[a_t^2\right] = \int_\R f(x) d\Q_{S_t}(x)$ requires computing the quantities $\int b^\Gamma_{i,n}(x) d\Q_{S_t}(x)$. In the general case of an arbitrary risk-neutral distribution $\Q_{S_t}$, a numerical quadrature must be performed. However, in Section \ref{sec:arbitrage_free_interpolation}, we propose an interpolation method for the implied volatility surface which produces marginal distributions of the form $\Q_{S_t} = g \Q^0_t$ where the ``prior'' $\Q^0$ is the distribution of a simple base model like the Black-Scholes model and $g$ is a B-spline parameterization of the Radon-Nikodym derivative of $\Q_{S_t}$ with respect to $\Q^0_t$.
\par In this case, computing $\int b^\Gamma_{i,n}(x) d\Q_{S_t}(x) = \int b^\Gamma_{i,n}(x) g(x) d\Q^0_t(x)$ amounts to integrating a piecewise polynomial with respect to the $\Q^0_t$ measure, which can be exactly evaluated in the cases of a Black-Scholes or Bachelier prior. 
\subsection{Numerical experiments with the shape-constrained particle method}\label{sec:particle_method_numerical}
\par Let us consider a model where the stochastic volatility process is the exponential of an Ornstein-Uhlenbeck process (like in Scott's model \cite{ScottStochasticVolatility87}), and the Black-Scholes implied volatility surface to be recovered is flat. More precisely for a volatility of volatility $\nu$, a mean-reversion parameter $\theta$, we consider the following model:
\begin{equation}\label{eq:model_slv1}
\begin{array}{l}
\qquad \qquad \left\{\begin{array}{ll}
d S_t = l(t, S_t) a_t dW_t,\\
a_t = a_0 \exp\left( U_t \right),\end{array}\right.\\[3mm]
\textnormal{where} \ U_t \ \textnormal{is a centered Ornstein-Uhlenbeck process,} \\
dU_t = -\theta U_t dt + \nu dW_t^\sigma \quad \textnormal{and} \quad d \langle W, W^\sigma \rangle_t = \rho dt. 
\end{array}
\end{equation}
\par \noindent As in the previous section, we use the normal convention for instantaneous volatilities rather than the lognormal convention. The Black-Scholes implied volatility being flat, $\Q_{S_t}$ is lognormal and we have  $\sigma_{\textnormal{Dup}}(t,x) = \sigma_{BS} x$. In this case, we get $\E[a_t^2] = a_0^2 \exp\left(\frac{\nu^2}{\theta}\left(1 - e^{-2 \theta t}\right)\right)$.
\begin{remark}[Exact time stepping scheme for the volatility]
Using that the volatility is the exponential of an Ornstein-Uhlenbeck $U_t$ correlated with the spot's Brownian motion $W_t$, we can exactly simulate the bivariate Gaussian random variable $\left(W_{t_{i+1}}, U_{t_{i+1}}\right)$. Indeed, $U_{t_{i+1}} = U_{t_i} e^{-\theta (t_{i+1} - t_i)} + \nu \int_{t_i}^{t_{i+1}} e^{\theta (u - t_{i+1})} dW^{\sigma}_u$, and the centered Gaussian random variable $\Big(\int_{t_i}^{t_{i+1}} e^{\theta (u - t_{i+1})} dW^{\sigma}_u, W_{t_{i+1}} - W_{t_i} \Big)$ has covariance matrix $\left(\begin{array}{cc} \frac{1 - e^{-2\theta \left(t_{i+1}-t_i \right)}}{2\theta} & \rho \frac{1 - e^{-\theta \left(t_{i+1}-t_i \right)}}{\theta}\\ \rho \frac{1 - e^{-\theta \left(t_{i+1}-t_i \right)}}{\theta} & t_{i+1} - t_i \end{array}\right)$. 
\end{remark}
\vspace{2mm}
\par
\begin{figure}[!ht]
	\begin{minipage}[c]{0.45\linewidth}
	\begin{center}\textbf{Unconstrained}\end{center}
	\end{minipage} \hfill
	\begin{minipage}[c]{0.45\linewidth}
	\begin{center}\textbf{Constrained}\end{center}
	\end{minipage}\\
	\begin{minipage}[c]{0.45\linewidth}
	\includegraphics[height=2.1cm]{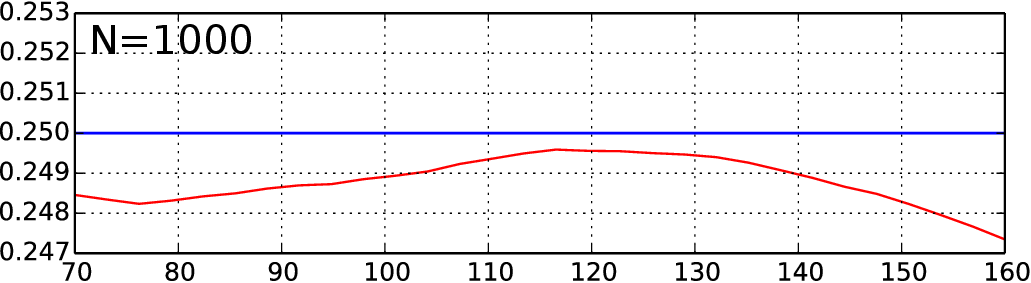}
	\end{minipage} \hfill
	\begin{minipage}[c]{0.45\linewidth}
	\includegraphics[height=2.1cm]{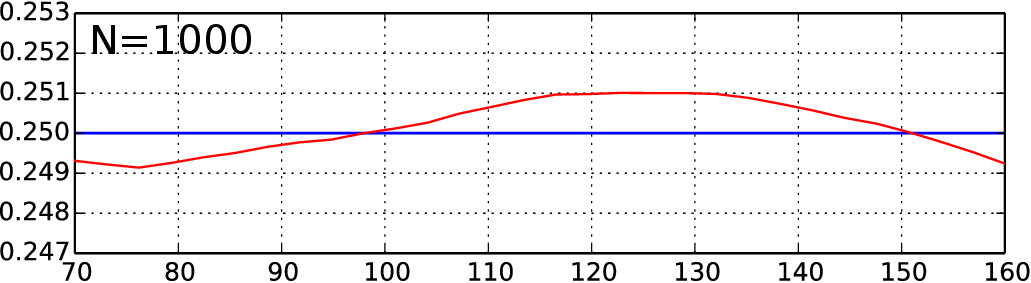}
	\end{minipage}\\
	\begin{minipage}[c]{0.45\linewidth}
	\includegraphics[height=2.1cm]{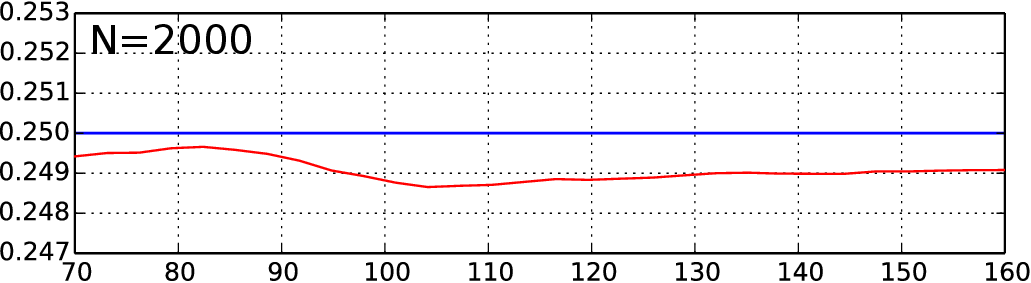}
	\end{minipage} \hfill
	\begin{minipage}[c]{0.45\linewidth}
	\includegraphics[height=2.1cm]{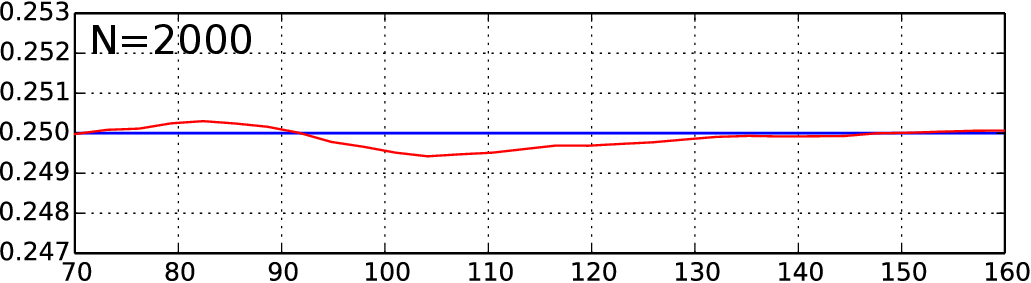}
	\end{minipage}\\
	\begin{minipage}[c]{0.45\linewidth}
	\includegraphics[height=2.1cm]{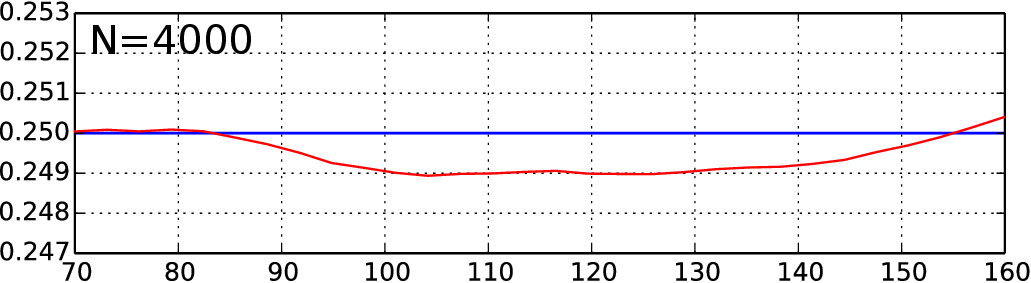}
	\end{minipage} \hfill
	\begin{minipage}[c]{0.45\linewidth}
	\includegraphics[height=2.1cm]{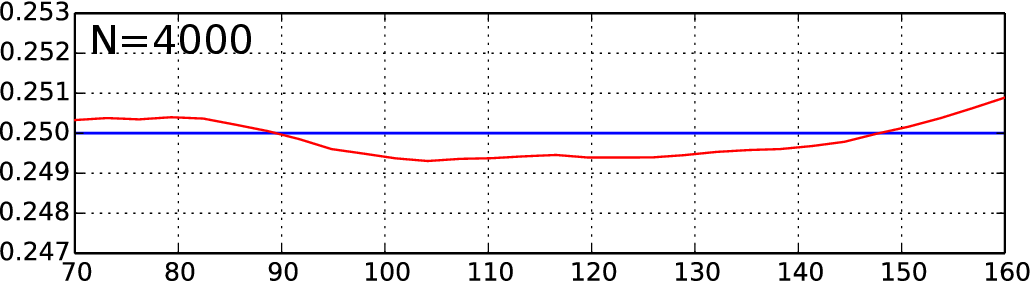}
	\end{minipage}\\
	\begin{minipage}[c]{0.45\linewidth}
	\includegraphics[height=2.1cm]{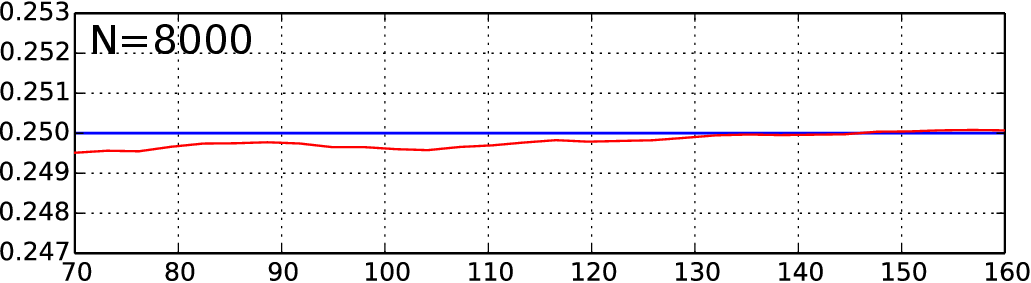}
	\end{minipage} \hfill
	\begin{minipage}[c]{0.45\linewidth}
	\includegraphics[height=2.1cm]{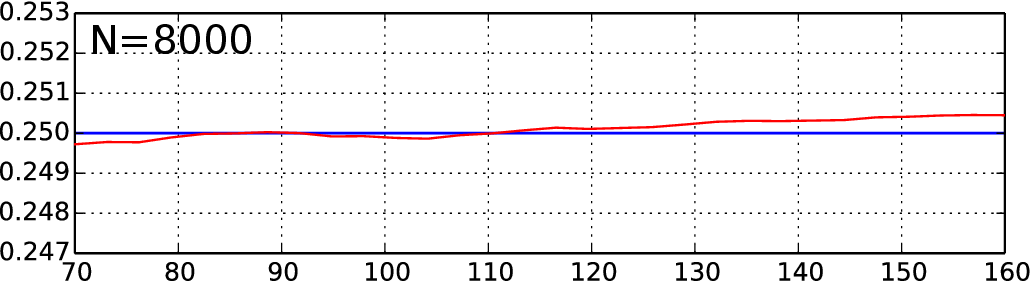}
	\end{minipage}\\
	\begin{minipage}[c]{0.45\linewidth}
	\includegraphics[height=2.1cm]{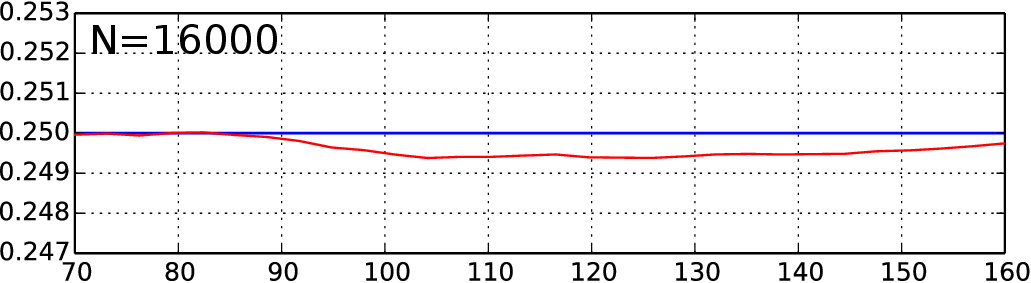}
	\end{minipage} \hfill
	\begin{minipage}[c]{0.45\linewidth}
	\includegraphics[height=2.1cm]{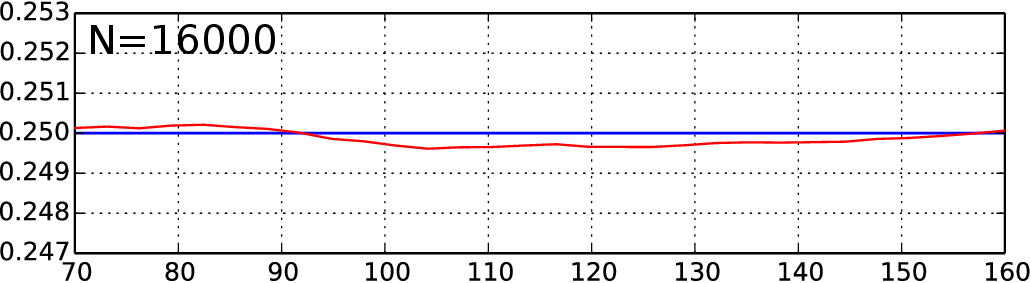}
	\end{minipage}	
	\caption{Implied volatility smiles of maturity $T=1$ in Model \eqref{eq:model_slv1}, with parameters $S_0=100$, $a_0=0.2$, $\theta = 1$, $\nu = 30\%$, $\rho = -0.8$. The leverage function $l(t,x)$ is calibrated so as to recover a flat Black-Scholes implied volatility smile $\sigma_{BS} = 25\%$. The size of the Monte Carlo sample used in the calibration is indicated in the top-left corner of each figure. On the left-hand side, the B-spline regressions involved in the particle method are constrained to be non-negative. On the right-hand side, regressions are also required to satisfy the equality constraint \eqref{eq:compability:expectation}. In both cases, we use $20$ knots, evenly spaced between $-2.5$ and $2.5$ standard deviations around the forward price in the log space. The regression uses Tikhonov regularization for the second derivative ($p=2$). The Tikhonov penalization factor is, $\sigma_{S_t}^{2p-1} / N$, as suggested in the previous section (Equation \eqref{eq:final_penalization_factor}).} 
	\label{fig:recover_BS}
\end{figure}

 In Figure \ref{fig:recover_BS}, we present the compared results of the particle method with and without taking into account the equality shape constrained \eqref{eq:compability:expectation} in the B-spline regression, with various sizes of the Monte Carlo sample. We observe that the equality constraint roughly results into a parallel shift in the eventual implied volatility smile, correcting the general bias of the Monte Carlo sample. In all cases, the B-spline regression is constrained to be non-negative (convex hull constraint \eqref{eq:compability:convexhull}). 
\par \noindent In our numerical experiments, the quadratic inequality constraint \eqref{eq:quadratic_constraint_slv} was inactive, and therefore it does not impact the results in practice, except in cases specially tailored so as to activate this constraint. 
\begin{remark}[Application to the calibration of local correlation models]
Another problem to which the particle method was successfully applied is the calibration of local correlation models (see e.g. \cite{GuyonAplusB}). Like in SLV models, the calibration procedure can benefit from shape constraints in the regressions (\eqref{eq:compability:convexhull} and \eqref{eq:compability:jensen}), and the other benefits of Tikhonov-regularized B-spline regression, but the equality constrained \eqref{eq:compability:expectation} cannot be enforced as we do not have an explicit form for the expectation of the regressed random variable.
\end{remark}

\section{Application to arbitrage-free completion of sparse option data}\label{sec:arbitrage_free_interpolation}
\par In this section, we address the problem of arbitrage-free completion of the vanilla option price surface from a sparse grid of vanilla option prices of various maturities and strikes. For the sake of simplicity, we first consider the case of a single maturity. 
\par We propose to use a B-spline parameterization of the Radon-Nikodym derivative of the risk-neutral distribution with respect to a simple roughly calibrated base model. The base model can be normal (Bachelier), lognormal (Black and Scholes \cite{BlackScholes73}) or more sophisticated like S.V.I. (Gatheral and Jacquier \cite{GatheralJacquierGlobalSVI}). The resulting calibrated risk-neutral density inherits certain properties of the base model. For example, in the case of the S.V.I. prior, the calibrated density will also have fat tails. 
\subsubsection*{Arbitrage conditions on vanilla option prices}
\par As we have seen in Section \ref{sec:shape_constraints}, convexity, integral constraints, equalities and inequalities on values and derivatives of B-splines qualify as linear constraints. All the conditions of absence of arbitrage on call option prices fall into these categories. Namely, if $K \mapsto C_T(K)$ and $K \mapsto P_T(K)$ are the (undiscounted) call and put option prices of strike $K$ and fixed maturity $T$ on some given underlying, the conditions of absence of arbitrage for a given maturity are given by:
\vspace{2mm}
\par \noindent \begin{minipage}[c]{0.48\linewidth}
\par For call option prices:
\begin{itemize}
\item $K \mapsto C_T(K)$ is nonnegative, nonincreasing and convex on $\R$,
\item $\lim\limits_{K \to +\infty} C_T(K) = 0$ and $\lim\limits_{K \to -\infty} C_T'(K) = -1$.
\end{itemize}
\end{minipage} \hfill
\begin{minipage}[c]{0.47\linewidth}
\par For put option prices:
\begin{itemize}
\item $K \mapsto P_T(K)$ is nonnegative, nondecreasing and convex on $\R$,
\item $\lim\limits_{K \to -\infty} P_T(K) = 0$ and $\lim\limits_{K \to +\infty} P_T'(K) = 1$.
\end{itemize}
\end{minipage}
\vspace{2mm}
\par All these conditions put together mean that the second derivative of the (undiscounted) call or put option price as a function of the strike in the sense of distributions is a probability distribution. This probability distribution is the risk-neutral probability of the underlying asset at maturity\footnote{Actually, it is the so-called forward-probability of the underlying asset, which coincides with the risk-neutral density if interest rates are assumed to be deterministic. We will not make the distinction in the rest of the paper.}. Arbitrage-free interpolation of vanilla option prices amounts to a density estimation problem. 
\par The forward price $F_T$ (the risk-neutral expectation of the underlying) must be equal to $\lim\limits_{K \to -\infty} C_T(K) + K$ and to $\lim\limits_{K \to +\infty} K - P_T(K)$. This amounts to $C_T(0) = F_T$ in the case where the underlying can only be nonnegative (such as a stock price). Bid-ask spreads on the forward price and call option prices, as well as firm equality constraints on these quantities qualify as linear constraints. Hence, there is a temptation to use a B-spline parameterization of the call option price as a function of strike (as in \cite{LauriniSplines} and \cite{FenglerSmoothing}). However, the resulting probability distribution would have compact support which is unrealistic. It has also been proposed to directly use a spline parameterization of the risk-neutral probability distribution \cite{MonteiroSplinePDF}, which has an identical flaw: the resulting density has compact support. 
\subsubsection*{B-spline parameterization of the Radon-Nikodym derivative with respect to a base model}
\par The method of our choice is to start from a prior ($L^1$) probability distribution, $\Q^0_{S_T}$. For example, $\Q^0_{S_T}$ can correspond to a rough calibration of the Black-Scholes model. We then use a B-spline parameterization of the Radon-Nikodym derivative of the risk-neutral distribution of $S_T$, $\Q_{S_T}$ with respect to $\Q^0_{S_T}$. For the sake of brevity, we will use the shorter notation $\Q^0_T:=\Q^0_{S_T}$ and $\Q_T:=\Q_{S_T}$.
$$
\Q_T = f_w \Q^0_T = \left(\sum\limits_{i = 0}^{k + n} w_i b^\Gamma_{i,n}\right) \Q^0_T.
$$
We truncate the B-spline to $0$th order (constant) extrapolation so that the resulting density is also ensured to be $L^1$. Constraints on $f_w$ are
$$
f_w \geq 0 \qquad \textnormal{and} \qquad \int_\R f_w(x) d\Q^0_T(x) = 1. 
$$
\par \noindent The forward price is given by 
$$
F_T = \int_\R x f_w(x) d\Q^0_T(x),
$$
while the undiscounted call and put option prices of strike $K$ are given by
$$
C_T(K) = \int_K^{+\infty} (x - K) f_w(x) d \Q^0_T(x) \qquad \textnormal{and} \qquad P_T(K) = \int_{-\infty}^K (K - x) f_w(x) d \Q^0_T(x).
$$
All these quantities happen to be linear forms of the loadings $w$. As a result, problems such as 
\begin{itemize}
\item the least-square fit of (mid) vanilla option prices under these constraints and a firm equality constraint on the forward,
\item minimization of the mean-square second derivative of $f_w$ under the firm constraint of yielding a price within bid-ask for each listed strike and for the forward price,
\end{itemize}
all qualify as second-order cone programs. 
\subsubsection*{Choice of the base model}
\par The advantage of lognormal base models is that integrals of the form $\int_a^b x^n d\Q^0_T(x)$ have closed-form expressions. However, this base model fails to account for implied volatility skew. The consequence is that the calibrated Radon-Nikodym density typically either explodes or almost vanishes in the wings and therefore, it is not very well approximated by piecewise polynomials. The advantage of more sophisticated models such as the SVI parameterization is that it can account for general features of the implied risk-neutral density of $S_T$ and thus more regular results for the Radon-Nikodym derivative. However, there is no closed-form expression for quantities of the form $\int_a^b x^n d\Q^0_T(x)$ if $n>1$, and one has to resort to approximate quadrature methods. 
\par
\begin{figure}[!ht]
	\begin{minipage}[c]{0.46\linewidth}
	\includegraphics[height=6.5cm]{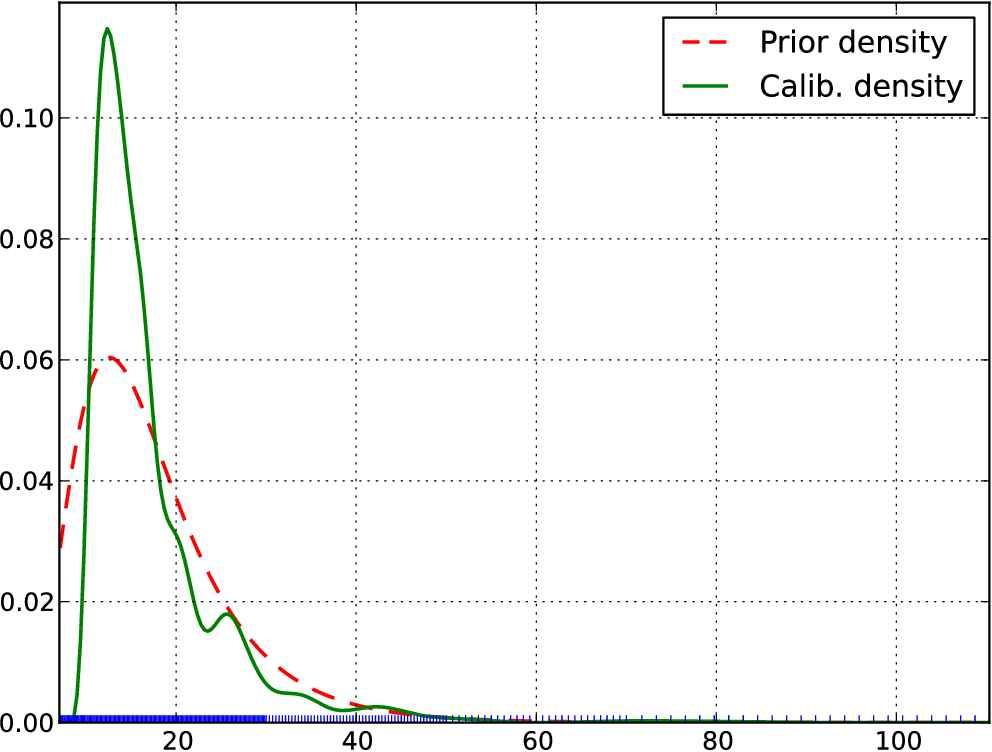}
	\end{minipage} \hfill
	\begin{minipage}[c]{0.46\linewidth}
	\includegraphics[height=6.5cm]{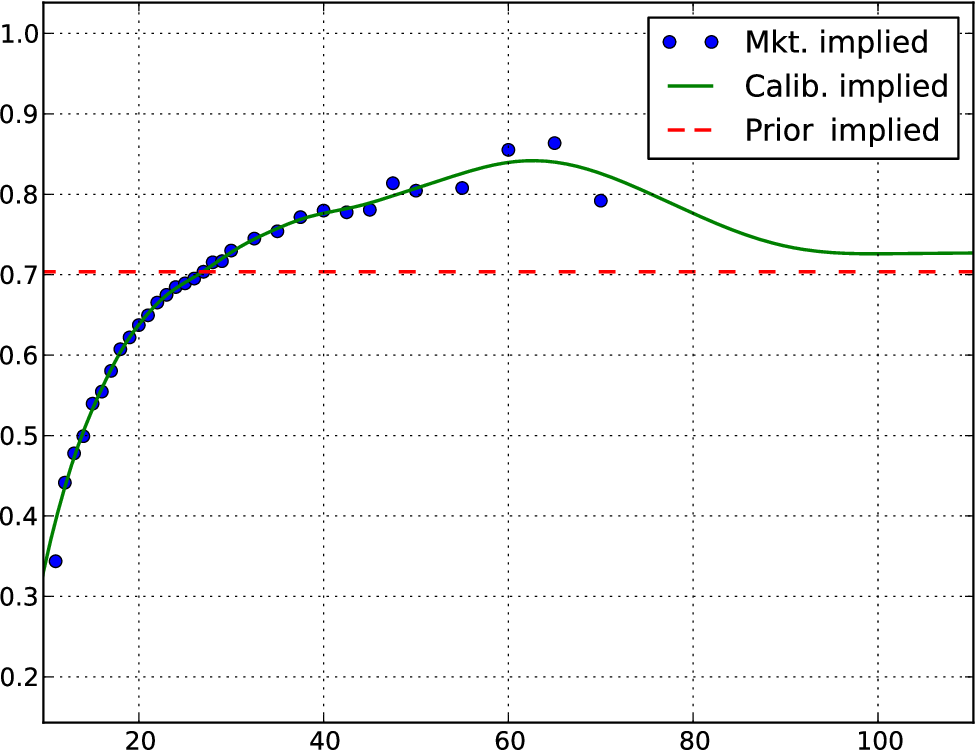}
	\end{minipage}
	\caption{Risk-neutral density (left) and implied volatility smile (right) of the VIX index for expiry October 16th, 2013 as seen on May 6th, 2013. The red dotted curve corresponds to the Black-Scholes base model, and the green curve to the calibrated implied risk-neutral density.}
	\label{fig:VIX_calibration}
\end{figure}

 In Figure \ref{fig:VIX_calibration}, we have calibrated a B-spline parameterization of the Radon-Nikodym derivative $\frac{d \Q_T}{d \Q^0_T}$ where $\Q_T$ is the risk-neutral probability distribution of the VIX volatility index on October 16th, 2013 seen from May 6th, 2013. 
\par
\begin{figure}[!ht]
	\begin{minipage}[c]{0.46\linewidth}
	\includegraphics[height=6.5cm]{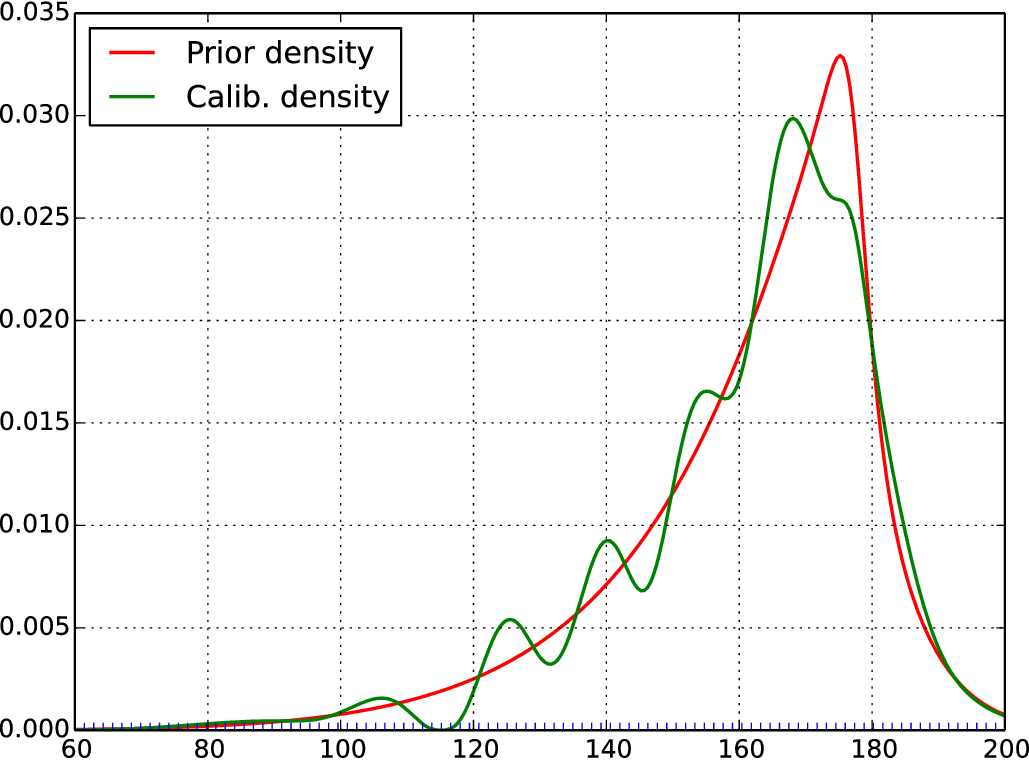}
	\end{minipage} \hfill
	\begin{minipage}[c]{0.46\linewidth}
	\includegraphics[height=6.5cm]{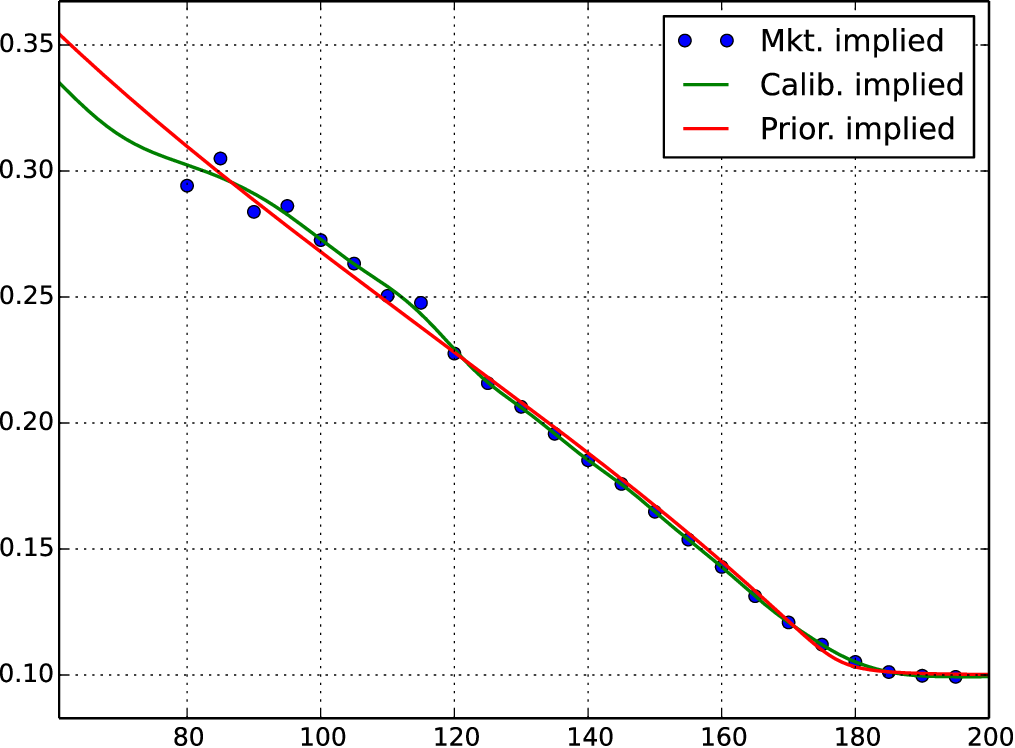}
	\end{minipage}
	\caption{Risk-neutral density (left) and implied volatility smile (right) of the Dow Jones index for expiry December 20th, 2014 as seen on April 25th, 2014. The red dotted curve corresponds to the roughly calibrated SVI base model, and the green curve to the calibrated implied risk-neutral density. The SVI parameters are $a=6.52e-3, b=0.036, \rho=-0.99, m = 0.091$ and $\sigma_0=0.027$.} 
	\label{fig:INDU_calibration}
\end{figure}

 In Figure \ref{fig:INDU_calibration}, we have calibrated a B-spline parameterization of the Radon-Nikodym derivative $\frac{d \Q_T}{d \Q^0_T}$ where $\Q_T$ is the risk-neutral probability distribution of the Dow Jones index for maturity December 20th, 2014 as seen on April 25th, 2014 around 11am. 
\par Eventually, we obtain a parameterization of the Radon-Nikodym density that also gives a smooth implied volatility smile. The volatility smile resulting from the procedure has the same asymptotics as the ones of the prior model, as the parameterized Radon-Nikodym derivative has a flat extrapolation. The prior model is typically parametric like Black-Scholes or the SVI parameterization which is flexible and often results in a decent fit. However, SVI cannot account for potential local features of the implied risk-neutral distribution such as bimodality. Our approach improves the fit of the parametric method while preserving the asymptotics and absence of arbitrage. 
\subsubsection*{Calendar arbitrage conditions}
\par So far, we have only addressed the case of a single maturity. If one assumes that rates are deterministic, the calendar arbitrage constraint is that undiscounted call and put option prices of constant relative strike (with respect to the forward price of the same maturity) is a non-decreasing function of maturity. 
\par
\begin{figure}
\begin{center}
\includegraphics[height=5.5cm]{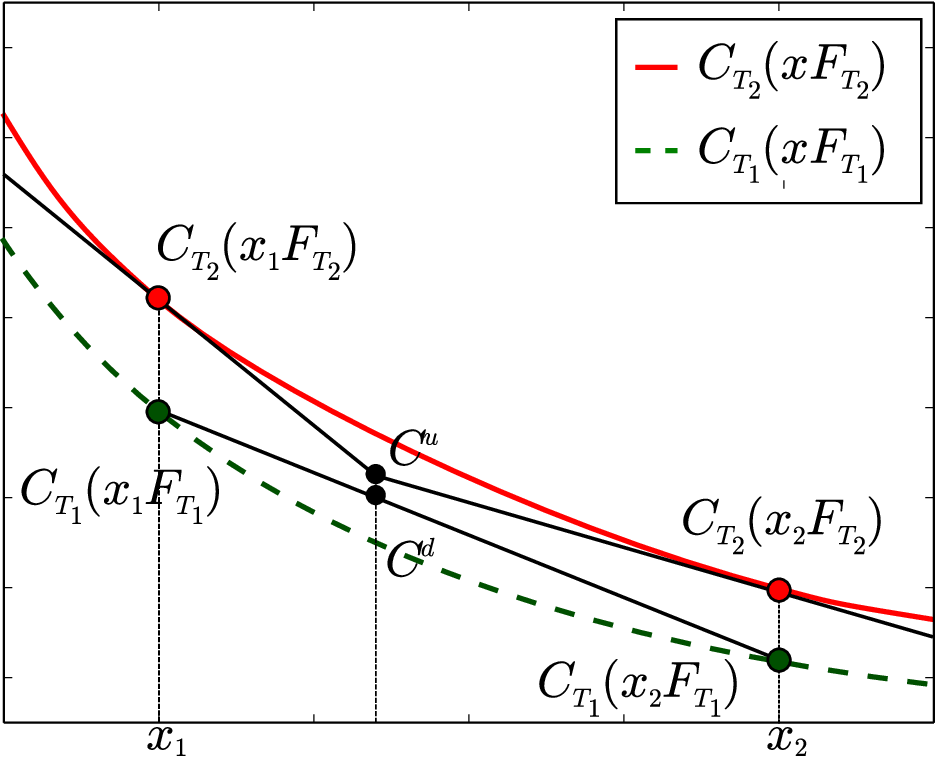}
\caption{$C^ u$ is the ordinate of the interpolation between the tangents of the call option price of maturity $T_2$ as a function of the relative strike $x$. $C^d$ is the value of the linear interpolation between $C_{T_1}(x_1 F_{T_1})$ and $C_{T_1}(x_2 F_{T_1})$ at the corresponding abscissa. $C^u \geq C^d$ is a sufficient condition for the absence of calendar arbitrage on $[x_1, x_2]$. The same equivalent argument holds for the Put option prices.}
\label{fig:sufficient_condition}
\end{center}
\end{figure}

 If we are given a sequence of listed maturities, we can enforce this condition on a fine grid of relative strikes, which amounts to a finite set of linear constraints. The arbitrage-free calibration of the implied volatility surface then amounts to a second-order cone program. However, this is not entirely satisfactory:
\begin{itemize}
\item On the one hand, we need the calendar arbitrage condition to be satisfied at any point lower than the first knot or higher than the last knot. If $0 \leq T_1 \leq T_2$ are two maturities and $x$ is a fixed relative strike, we denote $K_1 := x F_{T_1}$ and $K_2 := x F_{T_2}$. Now, if $x$ is such that $K_1$ and $K_2$ are both greater than the maximum knot, we have
$$
C_{T_1}(K_1) = \int_{K_1}^\infty w_{T_1} (x - K_1) d\Q^0_{T_1}(x) \qquad \textnormal{and} \qquad C_{T_2}(K_2) = \int_{K_2}^\infty w_{T_2} (x - K_2) d\Q^0_{T_2}(x),
$$
where $w_{T_1}$ and $w_{T_2}$ are the coefficients of the order-zero extrapolating B-spline basis function in the parameterizations of the Radon-Nikodym derivative for maturity $T_1$ and $T_2$ respectively. If the base model is assumed to be free of static arbitrage, $w_{T_2} \geq w_{T_1}$ is a sufficient condition to ensure that $C_{T_2}(K_2) \geq C_{T_1}(K_1)$. Regarding the low-strike extrapolation, the same argument holds when working with the put option prices. 
\item On the other hand, we need to ensure absence of calendar arbitrage in the interpolation region. In practice, we can settle for enforcing inequalities $C_{T_1}(x F_{T_1}) \leq C_{T_2}(x F_{T_2})$ only on a fine grid of values of $x$, for each couple of subsequent listed maturities $(T_1 < T_2)$. However, it can still happen that the calendar arbitrage occurs between two points of the grid. A sufficient condition to ensure absence of calendar arbitrage in the interpolation is detailed in Figure \ref{fig:sufficient_condition}. This second-order inequality condition is rather strong and it is the only nonlinear constraint. Therefore, for the numerical experiments, we settled for the fine grid condition. 
\end{itemize}
\par This discussion shows that the calibration of the Radon-Nikodym density with respect to a prior model, under the constraints of absence of arbitrage amounts to a global second-order cone program. Second order cone programs can be solved in minimum running time using available software such as CVXOPT \cite{cvxopt}, Mosek \cite{mosek} or CPLEX \cite{ilogcplex}. If the base model is Black-Scholes, the pricing of call and put options in the calibrated model remains explicit. The risk-neutral density in the calibrated model also has a closed-form expression. Hence, the denominator in the local volatility stripping formula \eqref{eq:stripping_form} has a closed-form expression. 
\subsubsection*{Smoothness in the time direction}
\par In \cite{JesperVolInterpolation}, Andreasen and Huge proposed an interpolation and extrapolation method which is guaranteed to yield arbitrage-free surfaces. Nonetheless, the resulting implied volatility surface fails to be differentiable in time at listed market maturities and thus, the corresponding local volatility has discontinuities in the time direction. Even though it is not formally an arbitrage, one could take advantage of a market maker using this method in markets where listed maturities are sliding with the current date, like foreign exchange markets. If the discontinuity is positive one can sell a ``calendar butterfly'' $C_{T+h} + C_{T - h} - 2 C_T$ where $T$ is a given market maturity and $h$ is $1$ day. One day later, all the tenors of the corresponding options will be lower than the new listed market maturity and the calendar butterfly will be valued at (almost) $0$ by the market maker if he has recalibrated his model. One can then buy it back for a much lower price. In the case where the discontinuity is negative, we can apply the opposite strategy. In short, in option markets with sliding listing maturity, market makers re-calibrating their models every day expose themselves to "calendar butterfly" strategies if their calibration procedure for the implied volatility surface has discontinuities in the time direction. 
\par Therefore, one should, wherever possible, produce implied volatility surfaces that are $C^1$ in the time direction. This can be ensured in our method by adding a term in the regularity penalization corresponding to the second time-derivative of loadings of the B-spline basis functions.
\subsubsection*{Choice of the base model}
\par
\begin{figure}[!ht]
	\begin{minipage}[c]{0.46\linewidth}
	\includegraphics[height=6cm]{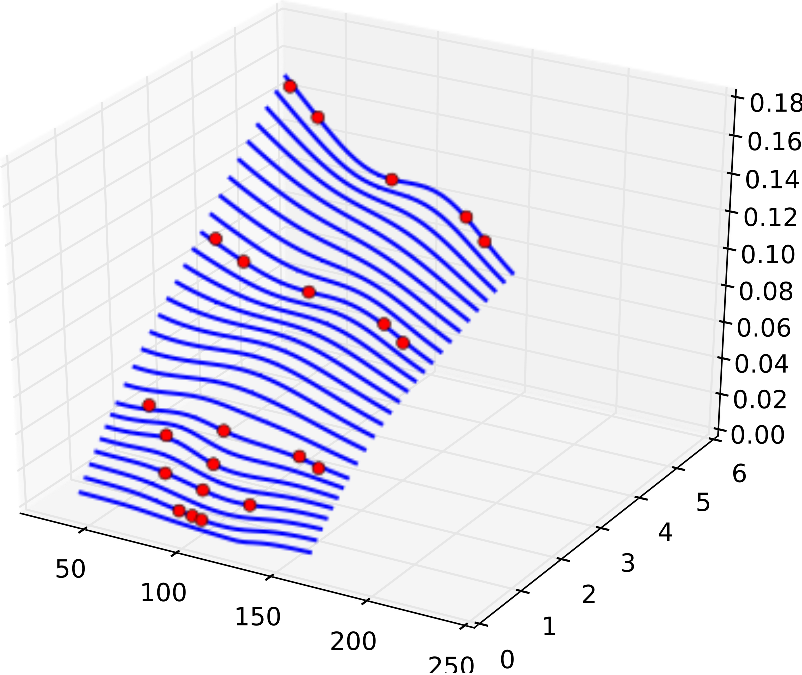}
	\end{minipage} \hfill
	\begin{minipage}[c]{0.46\linewidth}
	\includegraphics[height=6cm]{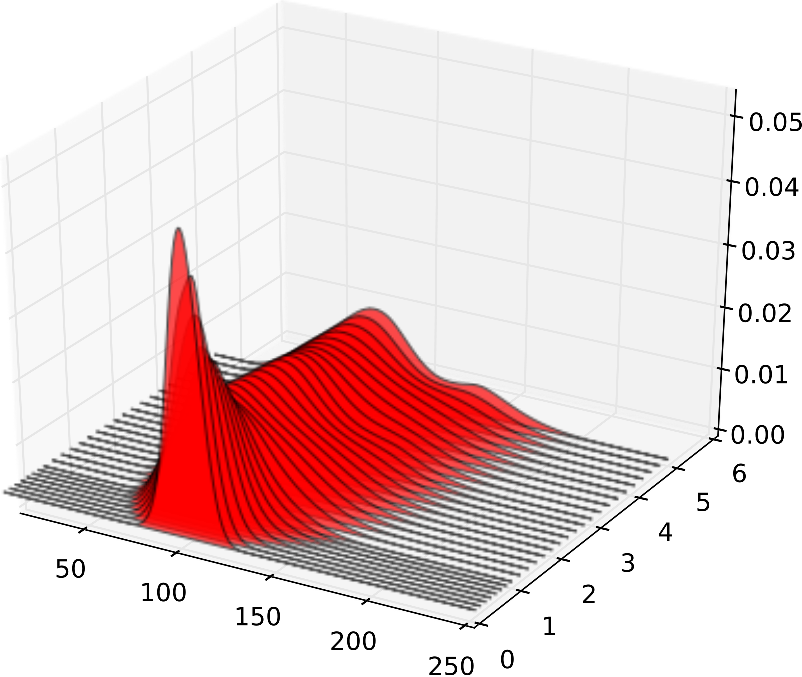}
	\end{minipage}
	\caption{Calibration of Radon-Nikodym densities from sparse vanilla option data. On the left-hand side, we plot the sequence of implied total lognormal variances (input data and interpolation). On the right-hand side, we show the calibrated risk-neutral densities for the same maturities.}
	\label{fig:global_calibration}
\end{figure}

 Like in the case of a single maturity slice, we may choose a simple parametric model like Bachelier or Black-Scholes as a base model $\Q_t^0$, because quantities of the form $\int_a^b x^n d\Q_t^0(x), \ (a,b) \in \R^2$ have closed-form expressions for any  maturity $t$. For example, in Figure \ref{fig:global_calibration}, we display the results of the global calibration to a sparse grid of vanilla option prices. Some of the maturities do not have any listed option prices. We can add as many of those intermediate maturities as necessary to obtain a fine grid in the time direction. The base model is the Black \& Scholes model with volatility equal to $20\%$. 
\par However, in the case where the market smile is skewed, the Radon-Nikodym derivatives $\frac{d\Q_t}{d\Q_t^0}$ will either explode or vanish far in or out of the money. Therefore, we need a parametric base model, flexible enough to account for general features of the volatility smile like fat-tailness and skewness, while remaining globally arbitrage-free. Such a parametric form was discovered by Gatheral and Jacquier \cite{GatheralJacquierGlobalSVI}. They devised a large family of parametric models, which amounts to a usual SVI parameterization at each maturity slice. Although, it is still unable to account for local features of the volatility smiles like multi-modality. In this context, our Radon-Nikodym parameterization method can be seen as a multiplicative residual fitting algorithm for the SVI model, which still guarantees absence of static arbitrage. 
\subsubsection*{The global SVI prior (SSVI \cite{GatheralJacquierGlobalSVI})}
\par We denote the log-moneyness by $k := \log(x/F_t)$, where $F_t$ is the forward price of maturity $t$. The implied Black-Scholes total variance is denoted by $w(t,k) := \sigma_{\textnormal{BS}}^2(t,x) t$. The Black-Scholes formula then gives $C_{BS}(k) = F_t \left( \mathcal{N}(d_+(t,k)) - e^k \mathcal{N}(d_-(t,k)) \right), \qquad \textnormal{where} \qquad d_{\pm} := -\frac{k}{\sqrt{w(t,k)}} \pm \frac{\sqrt{w(t,k)}}{2}$. If we define $g(t,k) := \left( 1 - \frac{k w'}{2w}\right)^2 - \frac{g'}{4} \left( \frac{1}{w} + \frac{1}{4} \right) + \frac{w''}{2}$, the risk-neutral density of the spot at maturity $t$ is 
$$
p(t,x) = \frac{1}{x}\frac{g(t,k)}{\sqrt{2 \pi w(t,k)}} \exp\left(-\frac{d_-(t,k)^2}{2}\right) \qquad (k = \log(x / F_t)).
$$
For a given maturity, Gatheral and Jacquier proposed to use the following parametric form
$$
w(t,k) = \Delta_t + \frac{\theta_t}{2} \left( 1 + \zeta_t \rho k + \sqrt{(\zeta_t k + \rho)^2 + (1-\rho^2)}\right),
$$
which coincides with a special case of the original SVI parameterization, up to a change of variable. Then $\zeta_t$ is assumed to be a function of $\theta_t$, that is $\zeta_t = \phi(\theta_t)$. It is proved that this so-called SSVI parameterization is free of butterfly arbitrage if for all $\theta > 0$,
$$
\theta \phi(\theta) (1 + |\rho|) < 4 \quad \textnormal{and} \quad \theta \phi(\theta)^2 (1 + |\rho|) \leq 4.
$$
The absence of calendar arbitrage is ensured by the following set of conditions
$$
\begin{array}{ll}
\partial_t \theta \geq 0, \ \partial_t \Delta \geq 0 \ \textnormal{and} \ \Delta_t \geq 0 \quad \textnormal{for all} \ t \geq 0,\\
0 \leq \partial_\theta(\theta \phi(\theta)) \leq \frac{1}{\rho^2} \left( 1 + \sqrt{1 - \rho^2}\right) \phi(\theta) \quad \textnormal{for all} \ \theta >0. 
\end{array}
$$
\subsubsection*{Our choice of parametric global SSVI}
\par We used the following functional forms for  $\Delta(t)$, $\theta(t)$ and $\phi(\theta)$:
$$
\Delta(t) = C t, \quad \phi(\theta) \equiv \eta \theta^{-\gamma} \quad \textnormal{and} \quad \theta(t) = Kt \quad \textnormal{with} \quad C, \eta, K >0, \ \gamma \in (0, 1)\ \textnormal{and} \ \rho \in (-1,1) . 
$$
Eventually, the global implied volatility surface is determined by $5$ parameters, $(C, K, \rho, \eta, \gamma)$. It is clearly not sufficient for an accurate calibration of the implied volatility surface but it can already account for its general shape. In \cite{GatheralJacquierGlobalSVI}, Gatheral and Jacquier start from this parametric form, and then achieve a better fit by relaxing the rigid link between different maturity slices. We don't need to do so because we only use this parametric shape as a base model in our spline-based Radon-Nikodym calibration procedure. 
\subsubsection*{Numerical experiments with SSVI global prior model}
\par
\begin{figure}[!ht]
	\begin{minipage}[c]{0.46\linewidth}
	\includegraphics[height=6.5cm]{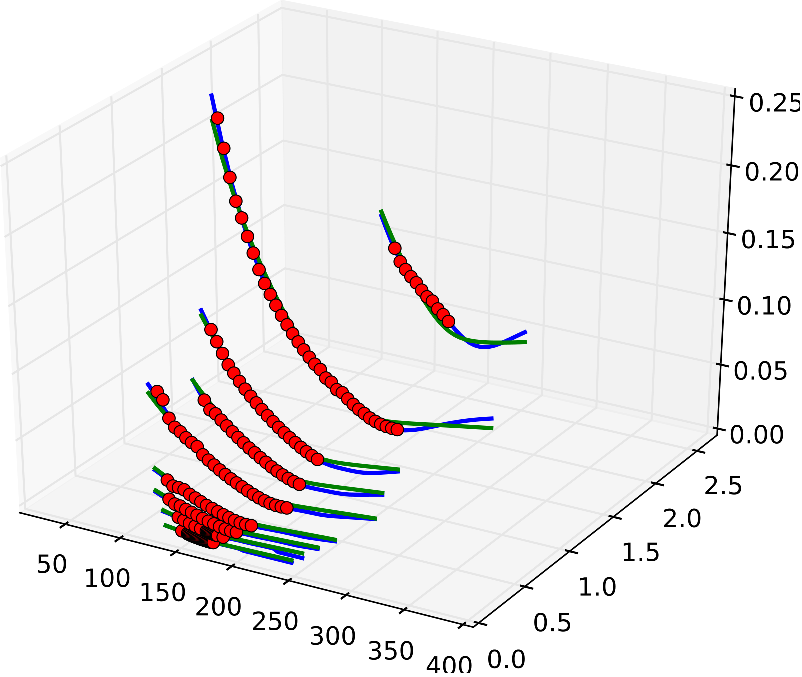}
	\end{minipage} \hfill
	\begin{minipage}[c]{0.46\linewidth}
	\includegraphics[height=6.5cm]{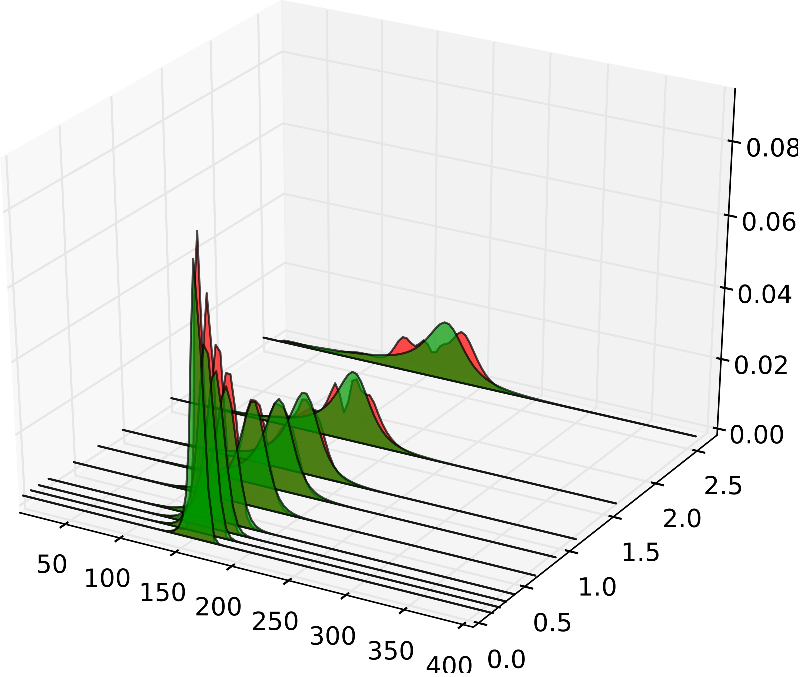}
	\end{minipage}
	\caption{Calibration of Radon-Nikodym densities from sparse vanilla option data on the Dow-Jones index. On the left-hand side, we plot the sequences of implied total lognormal variances (Input market data (red circles), interpolation (blue lines) and prior global SSVI fit (green lines)). On the right-hand side, we show the calibrated risk-neutral densities for the same maturities. The parameters for the roughly calibrated SSVI form are: $C = 0.0$, $\rho = -0.55$, $K = 0.019$, $\eta = 2.72$ and $\gamma = 0.25$.}
	\label{fig:INDU_calibration_full}
\end{figure}

 In Figure \ref{fig:INDU_calibration_full}, we display the results of the global calibration to the vanilla option prices on the Dow Jones index as of May, 28th 2014. 
\subsubsection*{Singular measures}
\par Even though the method proposed here relies on parameterizing the Radon-Nikodym derivatives of the marginals of the risk-neutral probability with respect to a prior model, it does not correspond to a global continuous change of measure. Indeed, if one calibrates (\textit{e.g.}) a local volatility model based on our interpolation, it will have a different volatility process from the prior model, while a continuous change of measure would only result in a change of drift. Therefore, the corresponding change of measure is singular. 
\section{Kolmogorov forward PDE for the Radon-Nikodym derivative}\label{sec:kolmo_radon_nikodym} 
\par As we have seen earlier, parameterizing the Radon-Nikodym derivative with respect to a simple base model allows to account for a prior belief on the risk-neutral distribution tails. If we believe in fat tails, we can use a fat-tailed prior model and the calibrated risk-neutral density will have the same rate of decay as the prior density. Moreover, as it is a linear transformation of the price space, conditions of absence of arbitrage remain linear conditions. 
\par In this section, we show that the Radon-Nikodym derivative is also a convenient space for the numerical treatment of the Kolmogorov forward PDE with the Galerkin method using B-spline finite elements. 
\par We consider the case of a simple S.D.E. $dS_t = \sqrt{v(t,S_t)} dW_t$ for some instantaneous variance $v(t,x)$ and deterministic initial condition $S_0$. The base model is assumed to be driven by the S.D.E. $dS_t = \sqrt{v_0(t,S_t)} dW_t$ where $v_0(0, S_0) = v(0, S_0)$. 
The Kolmogorov forward equation for the risk-neutral density $\phi(t, x)$ and the prior density $\phi_0$ are
$$
\partial_t \phi = \frac{1}{2} \partial^2_{xx} \left(v \phi\right) \qquad \textnormal{and} \qquad \partial_t \phi_0 = \frac{1}{2} \partial^2_{xx} \left(v_0 \phi_0 \right).
$$
\par \noindent If we denote by $f(t, x) := \frac{d\Q_t}{d\Q^0_t} = \frac{\phi(t,x)}{\phi_0(t, x)}$ the Radon-Nikodym derivative of the risk-neutral probability distribution of $S_t$ with respect to the base model, we obtain a linear PDE for $f$,
\begin{equation}\label{eq:radon_nikodym_pde}
\partial_t f = \frac{1}{2} \frac{\partial^2_{xx} (f v \phi_0)}{\phi_0} - f \frac{1}{2} \frac{\partial^2_{xx} (v_0 \phi_0)}{\phi_0}, \qquad \textnormal{and} \qquad f(0, x) \equiv 1. 
\end{equation}
\par The main benefit of this formulation with respect to the usual Kolmogorov forward PDE is that the initial condition is constant rather than being a Dirac mass. Hence, we do not need to tackle the numerical approximation of the Dirac mass with the basis functions. The formulation in terms of Radon-Nikodym derivative allows us to work with more regular functions. 
\par In the special case of a Bachelier base model (\textit{i.e.}, constant instantaneous variance $v_0$), $\phi_0(t,x) = \frac{1}{\sqrt{2 \pi v_0 t}} \exp\left( -\frac{(x - S_0)^2}{2 v_0 t} \right)$ and the terms $\frac{\partial_x (\phi_0)}{\phi_0}$ and $\frac{\partial_{xx} (\phi_0)}{\phi_0}$ involved in PDE \eqref{eq:radon_nikodym_pde} are polynomials of order $1$ and $2$ respectively. 
\subsubsection*{Variational formulation}
\par We first derive a spatially weak formulation of PDE \eqref{eq:radon_nikodym_pde}. Multiplying by a test function $u$ of the space variable $x$ and integrating over $\R$, we get
$$
\partial_t \int f u = \frac{1}{2} \int v \partial^2_{xx} (f) u + \int \left( \frac{\partial_x(v \phi_0)}{\phi_0}\right) \partial_x (f) u + \int \frac{1}{2} \left(\frac{\partial_{xx}^2((v - v_0)\phi_0)}{\phi_0}\right) f u.
$$
Using an integration by parts, we get the weak formulation of the PDE \begin{equation}\label{eq:weak_formulation}
\partial_t \int f u = -\frac{1}{2} \int v \partial_x (f) \partial_x (u) + \int \underbrace{\left( \frac{\partial_x( \phi_0)}{\phi_0} + \frac{1}{2} \partial_x (v) \right)}_{=: c} \partial_x (f) u + \int \underbrace{\frac{1}{2} \left(\frac{\partial_{xx}^2((v - v_0)\phi_0)}{\phi_0}\right)}_{=: e} f u.
\end{equation}
\subsubsection*{B-spline finite elements}
\par We plug a B-spline or order $n$, $f(t,x) = \sum\limits_{j = 0}^{k + n} w_j(t) b^\Gamma_{j, n}(x)$ into Equation \eqref{eq:weak_formulation} and we obtain
$$
\partial_t \sum\limits_{j = 0}^{k + n + 1} w_j(t) \int b^\Gamma_{j, n} u = -\frac{1}{2} \sum\limits_{j = 0}^{k + n + 1} w_j(t) \int v \partial_x \left(b^\Gamma_{j, n}\right) \partial_x (u) + \sum\limits_{j = 0}^{k + n + 1} w_j(t) \int c \partial_x \left(b^\Gamma_{j, n}\right) u + \sum\limits_{j = 0}^{k + n + 1} w_j(t)\int e b^\Gamma_{j, n} u,
$$
that is, $\partial_t \sum\limits_{j = 0}^{k + n + 1} w_j(t) \int b^\Gamma_{j, n} u = \sum\limits_{j = 0}^{k + n + 1} w_j(t) \left( -\frac{1}{2} \int v \partial_x \left(b^\Gamma_{j, n}\right) \partial_x (u) + \int c \partial_x \left(b^\Gamma_{j, n}\right) u + \int e b^\Gamma_{j, n} u \right)$.
We plug $u = b^\Gamma_{i, n}$ for $n < i < k$ (the basis functions of compact support) into \eqref{eq:weak_formulation}, we get a matrix equation $A \partial_t w = B(t) w$, where $A$ is defined by $A_{ij} = \int b^\Gamma_{j, n} b^\Gamma_{i, n}$, and is a band matrix independent of $t$, and $B$ is defined by
$$
B(t)_{ij} = -\frac{1}{2} \int v \partial_x \left(b^\Gamma_{j, n}\right) \partial_x (b^\Gamma_{i, n}) + \int c \partial_x \left(b^\Gamma_{j, n}\right) b^\Gamma_{i, n} + \int e b^\Gamma_{j, n} b^\Gamma_{i, n}, \qquad n < i < k,
$$ 
which is time-dependent (because of the time dependence of $v$, $c$ and $e$.)
If $p$ is the truncation order, we need to have as much as $2 (p + 1)$ additional linear equations or boundary conditions to obtain a full-rank system. For example, in the case of a flat extrapolation ($p = 0$), we need two boundary conditions. We can use the integral constraints
\begin{equation}\label{eq:integral_conditions}
\int_\R f(x) \phi_0(x)dx = 1 \qquad \textnormal{and} \qquad \int_\R x f(x) \phi_0 (x)dx = S_0.
\end{equation}
For higher order extrapolation, additional boundary conditions must be added to the system. In the end, we obtain a $(k + 2 p - n + 1)$-dimensional ordinary differential equation. This can be solved using an explicit or implicit time stepping scheme. 
\subsubsection*{Integral constraints, stability and singly bordered band diagonal linear systems}
\par In place of boundary conditions, the integral constraints \eqref{eq:integral_conditions} do not preserve the banded form of the system. The eventual system is \emph{singly bordered band diagonal}. Still, using a specialization of Gauss elimination to this specific case, we can take advantage of the sparsity with a $O(n m^2)$ complexity. We refer to \cite{GolubVanLoan} for more details on Gauss elimination for sparse matrices. 
\par Furthermore, in the stepping algorithm, rather than explicitly solving the linear system, one can use a least-square minimization under firm non-negativity constraints on the B-spline coefficients for the Radon-Nikodym derivative. In this setting, the time stepping is a quadratic program, which is slightly more expansive than solving the linear system. The upside is that together with the integral conditions \eqref{eq:integral_conditions}, the non-negativity constraint forces the result to be a valid Radon-Nikodym density, and therefore to have an $L^1(\PP_X)$ norm equal to $1$ at every time step. In this setting, both the explicit and implicit versions of the scheme are unconditionally stable. 
\subsubsection*{The time-independent case}
\par \noindent A PDE of the form $\partial_t f = \mathcal{L}_x f$ where $\mathcal{L}_x$ is a linear differential operator independent of $t$ can be seen as an infinite-dimensional ordinary differential equation whose solution is directly given by the exponential of the operator $\mathcal{L}_x$: $f(t, \cdot) \equiv \exp(\mathcal{L}_x) f_0(\cdot)$ where $f_0(\cdot) \equiv f(0, \cdot)$ is the initial condition. A space-discretized version of the PDE can be solved in the same way with a matrix exponential. This has been used for option pricing with time-independent local volatility models by Albanese and Trovato in \cite{AlbaneseTrovatoCMS}. However, even if the local variance $v$ is time-independent, the forward PDE for the Radon-Nikodym derivative \eqref{eq:radon_nikodym_pde} is not of the required form, unless we use a time-independent base density $\phi_0$. 
\section{Conclusion and future research}
\par In this article, we have applied techniques based on shape-constrained B-splines for two problems arising in volatility modeling.
\par The first one is the application to the recently developed particle method for the calibration the leverage function in SLV models. The regression problems arising in this case were typically tackled by non-parametric regression. Using shape-constrained B-splines, we could exploit additional information coming from the knowledge of the marginal distributions in addition to the bivariate sample. This led us to an improved accuracy with the same sample size.
\par The second problem we tackled is the arbitrage-free interpolation of the implied volatility surface. In the price space, arbitrage conditions are simply formulated but naive interpolation methods yield unrealistic implied volatility surfaces. Interpolating in the implied volatility space poses challenges as non-arbitrage constraints on implied volatility are non-linear. In this paper, we have shown that using a prior roughly calibrated model, a B-spline parameterization of the Radon-Nikodym density allows us to formulate the calibration problem as a second-order cone program and produces smooth implied volatility surfaces.
\par This article leaves some questions open. A first interesting problem is whether the necessary compatibility conditions \eqref{eq:compability:expectation}, \eqref{eq:compability:convexhull} and \eqref{eq:compability:jensen} between the marginals and the conditional expectation are sufficient to ensure the existence of a corresponding joint distribution. Another subject which has not been addressed in this paper is the problem of optimal knot selection for B-spline regression in the context of the particle method, as knots were generally evenly spaced on a certain range of standard deviations. A new approach to adaptive knot selection for multivariate B-spline is the subject of an ongoing research article by the author. 
\appendix
\section{The Abergel-Tachet PDE}\label{sec:abergel_tachet}
\par Let us consider the SDE \eqref{eq:slv_sde} in the case where the stochastic volatility process $a_t$ is itself a one-dimensional It\^o process. 
$$
\left\{\begin{array}{ll} dS_t = a_t l(t, S_t)dW_t,\\
da_t = \beta(t,a_t)dt + \nu(t,a_t) dW_t^\sigma,\\
d\langle W, W^\sigma \rangle_t = \rho dt.
\end{array}\right.
$$
\par \noindent Following Abergel and Tachet \cite{AbergelTachet} (except that we use normal conventions for volatilities rather than lognormal), the Kolmogorov forward PDE for the joint density $p$ of $(S_t, a_t)$ is
$$
\left\{\begin{array}{l}
\partial_t p = \frac{1}{2}\partial^2_{xx} ((l y)^2 p) + \rho \partial^2_{xy} ((ly) \nu p) + \frac{1}{2} \partial^2_{yy} (\nu^2 p) - \partial_y (\beta p),\\
p(0,x,y) = \delta_{S_0, a_0}(x,y).
\end{array}\right.
$$
\par \noindent Defining $q:=\int p dy$, we get $\partial_t q = \frac{1}{2} \partial^2_{xx} \left( l^2 \int y^2 p dy \right)$. Then, identifying the terms with the Dupire equation $\partial_t q = \frac{1}{2} \partial^2_{xx} \left( \sigma_{\textnormal{Dup}}^2 q \right)$, we recover the calibration condition \eqref{eq:calibration_condition}, that is, $l^2 = \frac{\sigma_{\textnormal{Dup}}^2 \int p dy}{\int y^2 p dy}$. The PDE corresponding to the McKean stochastic integro-differential equation \eqref{eq:nonlinear_sde} is then 
$$
\left\{\begin{array}{l}
\partial_t p = \frac{1}{2}\partial^2_{xx} \left(\frac{y^2 \int p dy}{\int y^2 p dy} \sigma_{\textnormal{Dup}}^2 p \right) + \rho \partial^2_{xy} \left(\frac{y \sqrt{\int p dy}}{\sqrt{\int y^2 p dy}} \sigma_{\textnormal{Dup}} \nu p\right) + \frac{1}{2} \partial^2_{yy} (\nu^2 p) - \partial_y (\beta p),\\
p(0,x,y) = \delta_{S_0, a_0}(x,y).
\end{array}\right.
$$

\vspace{6mm}
\par \noindent The author is grateful to Daniel Andor, Bruno Dupire, Julien Guyon, Alexey Polishchuk and Stephen Taylor for fruitful discussions and their remarks and comments on this work.
\bibliography{biblio}
\end{document}